\documentstyle[12pt]{article}
\newcommand {\cD}{{\cal D}}

\def\a{\alpha}
\def\b{\beta}
\def\c{\chi}
\def\d{\delta}
\def\e{\epsilon}

\def\k{\kappa}
\def\l{\lambda}
\def\m{\mu}
\def\n{\nu}

\def\q{\theta}
\def\r{\rho}
\def\s{\sigma}
\def\t{\tau}

\def\x{\xi}
\def\z{\zeta}

\def\L{\Lambda}

\def\P{\Pi}

\newcommand{\ad}{{\dot{\alpha}}}                           
\newcommand{\bd}{{\dot{\beta}}}                            

\newcommand{\hf}{\frac12}
\newcommand{\be}{\begin{equation}}
\newcommand{\ee}{\end{equation}}
\newcommand{\bea}{\begin{eqnarray}}
\newcommand{\eea}{\end{eqnarray}}

\def\covbox {\stackrel{\frown}{\Box}}

\def \N {{\cal N}}

\def\d{\partial}

\def \N {{\cal N }}

\begin{document}
\setcounter{page}{0}
\immediate\write16{<<WARNING: LINEDRAW macros work with emTeX-dvivers
                    and other drivers supporting emTeX \special's
                    (dviscr, dvihplj, dvidot, dvips, dviwin, etc.) >>}
\newdimen\Lengthunit       \Lengthunit  = 1.5cm
\newcount\Nhalfperiods     \Nhalfperiods= 9
\newcount\magnitude        \magnitude = 1000

\catcode`\*=11
\newdimen\L*   \newdimen\d*   \newdimen\d**
\newdimen\dm*  \newdimen\dd*  \newdimen\dt*
\newdimen\a*   \newdimen\b*   \newdimen\c*
\newdimen\a**  \newdimen\b**
\newdimen\xL*  \newdimen\yL*
\newdimen\rx*  \newdimen\ry*
\newdimen\tmp* \newdimen\linwid*

\newcount\k*   \newcount\l*   \newcount\m*
\newcount\k**  \newcount\l**  \newcount\m**
\newcount\n*   \newcount\dn*  \newcount\r*
\newcount\N*   \newcount\*one \newcount\*two  \*one=1 \*two=2
\newcount\*ths \*ths=1000
\newcount\angle*  \newcount\q*  \newcount\q**
\newcount\angle** \angle**=0
\newcount\sc*     \sc*=0

\newtoks\cos*  \cos*={1}
\newtoks\sin*  \sin*={0}

\catcode`\[=13

\def\rotate(#1){\advance\angle**#1\angle*=\angle**
\q**=\angle*\ifnum\q**<0\q**=-\q**\fi
\ifnum\q**>360\q*=\angle*\divide\q*360\multiply\q*360\advance\angle*-\q*\fi
\ifnum\angle*<0\advance\angle*360\fi\q**=\angle*\divide\q**90\q**=\q**
\def\sgcos*{+}\def\sgsin*{+}\relax
\ifcase\q**\or
 \def\sgcos*{-}\def\sgsin*{+}\or
 \def\sgcos*{-}\def\sgsin*{-}\or
 \def\sgcos*{+}\def\sgsin*{-}\else\fi
\q*=\q**
\multiply\q*90\advance\angle*-\q*
\ifnum\angle*>45\sc*=1\angle*=-\angle*\advance\angle*90\else\sc*=0\fi
\def[##1,##2]{\ifnum\sc*=0\relax
\edef\cs*{\sgcos*.##1}\edef\sn*{\sgsin*.##2}\ifcase\q**\or
 \edef\cs*{\sgcos*.##2}\edef\sn*{\sgsin*.##1}\or
 \edef\cs*{\sgcos*.##1}\edef\sn*{\sgsin*.##2}\or
 \edef\cs*{\sgcos*.##2}\edef\sn*{\sgsin*.##1}\else\fi\else
\edef\cs*{\sgcos*.##2}\edef\sn*{\sgsin*.##1}\ifcase\q**\or
 \edef\cs*{\sgcos*.##1}\edef\sn*{\sgsin*.##2}\or
 \edef\cs*{\sgcos*.##2}\edef\sn*{\sgsin*.##1}\or
 \edef\cs*{\sgcos*.##1}\edef\sn*{\sgsin*.##2}\else\fi\fi
\cos*={\cs*}\sin*={\sn*}\global\edef\gcos*{\cs*}\global\edef\gsin*{\sn*}}\relax
\ifcase\angle*[9999,0]\or
[999,017]\or[999,034]\or[998,052]\or[997,069]\or[996,087]\or
[994,104]\or[992,121]\or[990,139]\or[987,156]\or[984,173]\or
[981,190]\or[978,207]\or[974,224]\or[970,241]\or[965,258]\or
[961,275]\or[956,292]\or[951,309]\or[945,325]\or[939,342]\or
[933,358]\or[927,374]\or[920,390]\or[913,406]\or[906,422]\or
[898,438]\or[891,453]\or[882,469]\or[874,484]\or[866,499]\or
[857,515]\or[848,529]\or[838,544]\or[829,559]\or[819,573]\or
[809,587]\or[798,601]\or[788,615]\or[777,629]\or[766,642]\or
[754,656]\or[743,669]\or[731,681]\or[719,694]\or[707,707]\or
\else[9999,0]\fi}

\catcode`\[=12

\def\GRAPH(hsize=#1)#2{\hbox to #1\Lengthunit{#2\hss}}

\def\Linewidth#1{\global\linwid*=#1\relax
\global\divide\linwid*10\global\multiply\linwid*\mag
\global\divide\linwid*100\special{em:linewidth \the\linwid*}}

\Linewidth{.4pt}
\def\sm*{\special{em:moveto}}
\def\sl*{\special{em:lineto}}
\let\moveto=\sm*
\let\lineto=\sl*
\newbox\spm*   \newbox\spl*
\setbox\spm*\hbox{\sm*}
\setbox\spl*\hbox{\sl*}

\def\mov#1(#2,#3)#4{\rlap{\L*=#1\Lengthunit
\xL*=#2\L* \yL*=#3\L*
\xL*=\xscale\xL* \yL*=\yscale\yL*
\rx* \the\cos*\xL* \tmp* \the\sin*\yL* \advance\rx*-\tmp*
\ry* \the\cos*\yL* \tmp* \the\sin*\xL* \advance\ry*\tmp*
\kern\rx*\raise\ry*\hbox{#4}}}

\def\rmov*(#1,#2)#3{\rlap{\xL*=#1\yL*=#2\relax
\rx* \the\cos*\xL* \tmp* \the\sin*\yL* \advance\rx*-\tmp*
\ry* \the\cos*\yL* \tmp* \the\sin*\xL* \advance\ry*\tmp*
\kern\rx*\raise\ry*\hbox{#3}}}

\def\lin#1(#2,#3){\rlap{\sm*\mov#1(#2,#3){\sl*}}}

\def\arr*(#1,#2,#3){\rmov*(#1\dd*,#1\dt*){\sm*
\rmov*(#2\dd*,#2\dt*){\rmov*(#3\dt*,-#3\dd*){\sl*}}\sm*
\rmov*(#2\dd*,#2\dt*){\rmov*(-#3\dt*,#3\dd*){\sl*}}}}

\def\arrow#1(#2,#3){\rlap{\lin#1(#2,#3)\mov#1(#2,#3){\relax
\d**=-.012\Lengthunit\dd*=#2\d**\dt*=#3\d**
\arr*(1,10,4)\arr*(3,8,4)\arr*(4.8,4.2,3)}}}

\def\arrlin#1(#2,#3){\rlap{\L*=#1\Lengthunit\L*=.5\L*
\lin#1(#2,#3)\rmov*(#2\L*,#3\L*){\arrow.1(#2,#3)}}}

\def\dasharrow#1(#2,#3){\rlap{{\Lengthunit=0.9\Lengthunit
\dashlin#1(#2,#3)\mov#1(#2,#3){\sm*}}\mov#1(#2,#3){\sl*
\d**=-.012\Lengthunit\dd*=#2\d**\dt*=#3\d**
\arr*(1,10,4)\arr*(3,8,4)\arr*(4.8,4.2,3)}}}

\def\clap#1{\hbox to 0pt{\hss #1\hss}}

\def\ind(#1,#2)#3{\rlap{\L*=.1\Lengthunit
\xL*=#1\L* \yL*=#2\L*
\rx* \the\cos*\xL* \tmp* \the\sin*\yL* \advance\rx*-\tmp*
\ry* \the\cos*\yL* \tmp* \the\sin*\xL* \advance\ry*\tmp*
\kern\rx*\raise\ry*\hbox{\lower2pt\clap{$#3$}}}}

\def\sh*(#1,#2)#3{\rlap{\dm*=\the\n*\d**
\xL*=\xscale\dm* \yL*=\yscale\dm* \xL*=#1\xL* \yL*=#2\yL*
\rx* \the\cos*\xL* \tmp* \the\sin*\yL* \advance\rx*-\tmp*
\ry* \the\cos*\yL* \tmp* \the\sin*\xL* \advance\ry*\tmp*
\kern\rx*\raise\ry*\hbox{#3}}}

\def\calcnum*#1(#2,#3){\a*=1000sp\b*=1000sp\a*=#2\a*\b*=#3\b*
\ifdim\a*<0pt\a*-\a*\fi\ifdim\b*<0pt\b*-\b*\fi
\ifdim\a*>\b*\c*=.96\a*\advance\c*.4\b*
\else\c*=.96\b*\advance\c*.4\a*\fi
\k*\a*\multiply\k*\k*\l*\b*\multiply\l*\l*
\m*\k*\advance\m*\l*\n*\c*\r*\n*\multiply\n*\n*
\dn*\m*\advance\dn*-\n*\divide\dn*2\divide\dn*\r*
\advance\r*\dn*
\c*=\the\Nhalfperiods5sp\c*=#1\c*\ifdim\c*<0pt\c*-\c*\fi
\multiply\c*\r*\N*\c*\divide\N*10000}

\def\dashlin#1(#2,#3){\rlap{\calcnum*#1(#2,#3)\relax
\d**=#1\Lengthunit\ifdim\d**<0pt\d**-\d**\fi
\divide\N*2\multiply\N*2\advance\N*\*one
\divide\d**\N*\sm*\n*\*one\sh*(#2,#3){\sl*}\loop
\advance\n*\*one\sh*(#2,#3){\sm*}\advance\n*\*one
\sh*(#2,#3){\sl*}\ifnum\n*<\N*\repeat}}

\def\dashdotlin#1(#2,#3){\rlap{\calcnum*#1(#2,#3)\relax
\d**=#1\Lengthunit\ifdim\d**<0pt\d**-\d**\fi
\divide\N*2\multiply\N*2\advance\N*1\multiply\N*2\relax
\divide\d**\N*\sm*\n*\*two\sh*(#2,#3){\sl*}\loop
\advance\n*\*one\sh*(#2,#3){\kern-1.48pt\lower.5pt\hbox{\rm.}}\relax
\advance\n*\*one\sh*(#2,#3){\sm*}\advance\n*\*two
\sh*(#2,#3){\sl*}\ifnum\n*<\N*\repeat}}

\def\shl*(#1,#2)#3{\kern#1#3\lower#2#3\hbox{\unhcopy\spl*}}

\def\trianglin#1(#2,#3){\rlap{\toks0={#2}\toks1={#3}\calcnum*#1(#2,#3)\relax
\dd*=.57\Lengthunit\dd*=#1\dd*\divide\dd*\N*
\divide\dd*\*ths \multiply\dd*\magnitude
\d**=#1\Lengthunit\ifdim\d**<0pt\d**-\d**\fi
\multiply\N*2\divide\d**\N*\sm*\n*\*one\loop
\shl**{\dd*}\dd*-\dd*\advance\n*2\relax
\ifnum\n*<\N*\repeat\n*\N*\shl**{0pt}}}

\def\wavelin#1(#2,#3){\rlap{\toks0={#2}\toks1={#3}\calcnum*#1(#2,#3)\relax
\dd*=.23\Lengthunit\dd*=#1\dd*\divide\dd*\N*
\divide\dd*\*ths \multiply\dd*\magnitude
\d**=#1\Lengthunit\ifdim\d**<0pt\d**-\d**\fi
\multiply\N*4\divide\d**\N*\sm*\n*\*one\loop
\shl**{\dd*}\dt*=1.3\dd*\advance\n*\*one
\shl**{\dt*}\advance\n*\*one
\shl**{\dd*}\advance\n*\*two
\dd*-\dd*\ifnum\n*<\N*\repeat\n*\N*\shl**{0pt}}}

\def\w*lin(#1,#2){\rlap{\toks0={#1}\toks1={#2}\d**=\Lengthunit\dd*=-.12\d**
\divide\dd*\*ths \multiply\dd*\magnitude
\N*8\divide\d**\N*\sm*\n*\*one\loop
\shl**{\dd*}\dt*=1.3\dd*\advance\n*\*one
\shl**{\dt*}\advance\n*\*one
\shl**{\dd*}\advance\n*\*one
\shl**{0pt}\dd*-\dd*\advance\n*1\ifnum\n*<\N*\repeat}}

\def\l*arc(#1,#2)[#3][#4]{\rlap{\toks0={#1}\toks1={#2}\d**=\Lengthunit
\dd*=#3.037\d**\dd*=#4\dd*\dt*=#3.049\d**\dt*=#4\dt*\ifdim\d**>10mm\relax
\d**=.25\d**\n*\*one\shl**{-\dd*}\n*\*two\shl**{-\dt*}\n*3\relax
\shl**{-\dd*}\n*4\relax\shl**{0pt}\else
\ifdim\d**>5mm\d**=.5\d**\n*\*one\shl**{-\dt*}\n*\*two
\shl**{0pt}\else\n*\*one\shl**{0pt}\fi\fi}}

\def\d*arc(#1,#2)[#3][#4]{\rlap{\toks0={#1}\toks1={#2}\d**=\Lengthunit
\dd*=#3.037\d**\dd*=#4\dd*\d**=.25\d**\sm*\n*\*one\shl**{-\dd*}\relax
\n*3\relax\sh*(#1,#2){\xL*=\xscale\dd*\yL*=\yscale\dd*
\kern#2\xL*\lower#1\yL*\hbox{\sm*}}\n*4\relax\shl**{0pt}}}

\def\shl**#1{\c*=\the\n*\d**\d*=#1\relax
\a*=\the\toks0\c*\b*=\the\toks1\d*\advance\a*-\b*
\b*=\the\toks1\c*\d*=\the\toks0\d*\advance\b*\d*
\a*=\xscale\a*\b*=\yscale\b*
\rx* \the\cos*\a* \tmp* \the\sin*\b* \advance\rx*-\tmp*
\ry* \the\cos*\b* \tmp* \the\sin*\a* \advance\ry*\tmp*
\raise\ry*\rlap{\kern\rx*\unhcopy\spl*}}

\def\wlin*#1(#2,#3)[#4]{\rlap{\toks0={#2}\toks1={#3}\relax
\c*=#1\l*\c*\c*=.01\Lengthunit\m*\c*\divide\l*\m*
\c*=\the\Nhalfperiods5sp\multiply\c*\l*\N*\c*\divide\N*\*ths
\divide\N*2\multiply\N*2\advance\N*\*one
\dd*=.002\Lengthunit\dd*=#4\dd*\multiply\dd*\l*\divide\dd*\N*
\divide\dd*\*ths \multiply\dd*\magnitude
\d**=#1\multiply\N*4\divide\d**\N*\sm*\n*\*one\loop
\shl**{\dd*}\dt*=1.3\dd*\advance\n*\*one
\shl**{\dt*}\advance\n*\*one
\shl**{\dd*}\advance\n*\*two
\dd*-\dd*\ifnum\n*<\N*\repeat\n*\N*\shl**{0pt}}}

\def\wavebox#1{\setbox0\hbox{#1}\relax
\a*=\wd0\advance\a*14pt\b*=\ht0\advance\b*\dp0\advance\b*14pt\relax
\hbox{\kern9pt\relax
\rmov*(0pt,\ht0){\rmov*(-7pt,7pt){\wlin*\a*(1,0)[+]\wlin*\b*(0,-1)[-]}}\relax
\rmov*(\wd0,-\dp0){\rmov*(7pt,-7pt){\wlin*\a*(-1,0)[+]\wlin*\b*(0,1)[-]}}\relax
\box0\kern9pt}}

\def\rectangle#1(#2,#3){\relax
\lin#1(#2,0)\lin#1(0,#3)\mov#1(0,#3){\lin#1(#2,0)}\mov#1(#2,0){\lin#1(0,#3)}}

\def\dashrectangle#1(#2,#3){\dashlin#1(#2,0)\dashlin#1(0,#3)\relax
\mov#1(0,#3){\dashlin#1(#2,0)}\mov#1(#2,0){\dashlin#1(0,#3)}}

\def\waverectangle#1(#2,#3){\L*=#1\Lengthunit\a*=#2\L*\b*=#3\L*
\ifdim\a*<0pt\a*-\a*\def\x*{-1}\else\def\x*{1}\fi
\ifdim\b*<0pt\b*-\b*\def\y*{-1}\else\def\y*{1}\fi
\wlin*\a*(\x*,0)[-]\wlin*\b*(0,\y*)[+]\relax
\mov#1(0,#3){\wlin*\a*(\x*,0)[+]}\mov#1(#2,0){\wlin*\b*(0,\y*)[-]}}

\def\calcparab*{\ifnum\n*>\m*\k*\N*\advance\k*-\n*\else\k*\n*\fi
\a*=\the\k* sp\a*=10\a*\b*\dm*\advance\b*-\a*\k*\b*
\a*=\the\*ths\b*\divide\a*\l*\multiply\a*\k*
\divide\a*\l*\k*\*ths\r*\a*\advance\k*-\r*\dt*=\the\k*\L*}

\def\arcto#1(#2,#3)[#4]{\rlap{\toks0={#2}\toks1={#3}\calcnum*#1(#2,#3)\relax
\dm*=135sp\dm*=#1\dm*\d**=#1\Lengthunit\ifdim\dm*<0pt\dm*-\dm*\fi
\multiply\dm*\r*\a*=.3\dm*\a*=#4\a*\ifdim\a*<0pt\a*-\a*\fi
\advance\dm*\a*\N*\dm*\divide\N*10000\relax
\divide\N*2\multiply\N*2\advance\N*\*one
\L*=-.25\d**\L*=#4\L*\divide\d**\N*\divide\L*\*ths
\m*\N*\divide\m*2\dm*=\the\m*5sp\l*\dm*\sm*\n*\*one\loop
\calcparab*\shl**{-\dt*}\advance\n*1\ifnum\n*<\N*\repeat}}

\def\arrarcto#1(#2,#3)[#4]{\L*=#1\Lengthunit\L*=.54\L*
\arcto#1(#2,#3)[#4]\rmov*(#2\L*,#3\L*){\d*=.457\L*\d*=#4\d*\d**-\d*
\rmov*(#3\d**,#2\d*){\arrow.02(#2,#3)}}}

\def\dasharcto#1(#2,#3)[#4]{\rlap{\toks0={#2}\toks1={#3}\relax
\calcnum*#1(#2,#3)\dm*=\the\N*5sp\a*=.3\dm*\a*=#4\a*\ifdim\a*<0pt\a*-\a*\fi
\advance\dm*\a*\N*\dm*
\divide\N*20\multiply\N*2\advance\N*1\d**=#1\Lengthunit
\L*=-.25\d**\L*=#4\L*\divide\d**\N*\divide\L*\*ths
\m*\N*\divide\m*2\dm*=\the\m*5sp\l*\dm*
\sm*\n*\*one\loop\calcparab*
\shl**{-\dt*}\advance\n*1\ifnum\n*>\N*\else\calcparab*
\sh*(#2,#3){\xL*=#3\dt* \yL*=#2\dt*
\rx* \the\cos*\xL* \tmp* \the\sin*\yL* \advance\rx*\tmp*
\ry* \the\cos*\yL* \tmp* \the\sin*\xL* \advance\ry*-\tmp*
\kern\rx*\lower\ry*\hbox{\sm*}}\fi
\advance\n*1\ifnum\n*<\N*\repeat}}

\def\*shl*#1{\c*=\the\n*\d**\advance\c*#1\a**\d*\dt*\advance\d*#1\b**
\a*=\the\toks0\c*\b*=\the\toks1\d*\advance\a*-\b*
\b*=\the\toks1\c*\d*=\the\toks0\d*\advance\b*\d*
\rx* \the\cos*\a* \tmp* \the\sin*\b* \advance\rx*-\tmp*
\ry* \the\cos*\b* \tmp* \the\sin*\a* \advance\ry*\tmp*
\raise\ry*\rlap{\kern\rx*\unhcopy\spl*}}

\def\calcnormal*#1{\b**=10000sp\a**\b**\k*\n*\advance\k*-\m*
\multiply\a**\k*\divide\a**\m*\a**=#1\a**\ifdim\a**<0pt\a**-\a**\fi
\ifdim\a**>\b**\d*=.96\a**\advance\d*.4\b**
\else\d*=.96\b**\advance\d*.4\a**\fi
\d*=.01\d*\r*\d*\divide\a**\r*\divide\b**\r*
\ifnum\k*<0\a**-\a**\fi\d*=#1\d*\ifdim\d*<0pt\b**-\b**\fi
\k*\a**\a**=\the\k*\dd*\k*\b**\b**=\the\k*\dd*}

\def\wavearcto#1(#2,#3)[#4]{\rlap{\toks0={#2}\toks1={#3}\relax
\calcnum*#1(#2,#3)\c*=\the\N*5sp\a*=.4\c*\a*=#4\a*\ifdim\a*<0pt\a*-\a*\fi
\advance\c*\a*\N*\c*\divide\N*20\multiply\N*2\advance\N*-1\multiply\N*4\relax
\d**=#1\Lengthunit\dd*=.012\d**
\divide\dd*\*ths \multiply\dd*\magnitude
\ifdim\d**<0pt\d**-\d**\fi\L*=.25\d**
\divide\d**\N*\divide\dd*\N*\L*=#4\L*\divide\L*\*ths
\m*\N*\divide\m*2\dm*=\the\m*0sp\l*\dm*
\sm*\n*\*one\loop\calcnormal*{#4}\calcparab*
\*shl*{1}\advance\n*\*one\calcparab*
\*shl*{1.3}\advance\n*\*one\calcparab*
\*shl*{1}\advance\n*2\dd*-\dd*\ifnum\n*<\N*\repeat\n*\N*\shl**{0pt}}}

\def\triangarcto#1(#2,#3)[#4]{\rlap{\toks0={#2}\toks1={#3}\relax
\calcnum*#1(#2,#3)\c*=\the\N*5sp\a*=.4\c*\a*=#4\a*\ifdim\a*<0pt\a*-\a*\fi
\advance\c*\a*\N*\c*\divide\N*20\multiply\N*2\advance\N*-1\multiply\N*2\relax
\d**=#1\Lengthunit\dd*=.012\d**
\divide\dd*\*ths \multiply\dd*\magnitude
\ifdim\d**<0pt\d**-\d**\fi\L*=.25\d**
\divide\d**\N*\divide\dd*\N*\L*=#4\L*\divide\L*\*ths
\m*\N*\divide\m*2\dm*=\the\m*0sp\l*\dm*
\sm*\n*\*one\loop\calcnormal*{#4}\calcparab*
\*shl*{1}\advance\n*2\dd*-\dd*\ifnum\n*<\N*\repeat\n*\N*\shl**{0pt}}}

\def\hr*#1{\L*=\xscale\Lengthunit\ifnum
\angle**=0\clap{\vrule width#1\L* height.1pt}\else
\L*=#1\L*\L*=.5\L*\rmov*(-\L*,0pt){\sm*}\rmov*(\L*,0pt){\sl*}\fi}

\def\shade#1[#2]{\rlap{\Lengthunit=#1\Lengthunit
\special{em:linewidth .001pt}\relax
\mov(0,#2.05){\hr*{.994}}\mov(0,#2.1){\hr*{.980}}\relax
\mov(0,#2.15){\hr*{.953}}\mov(0,#2.2){\hr*{.916}}\relax
\mov(0,#2.25){\hr*{.867}}\mov(0,#2.3){\hr*{.798}}\relax
\mov(0,#2.35){\hr*{.715}}\mov(0,#2.4){\hr*{.603}}\relax
\mov(0,#2.45){\hr*{.435}}\special{em:linewidth \the\linwid*}}}

\def\dshade#1[#2]{\rlap{\special{em:linewidth .001pt}\relax
\Lengthunit=#1\Lengthunit\if#2-\def\t*{+}\else\def\t*{-}\fi
\mov(0,\t*.025){\relax
\mov(0,#2.05){\hr*{.995}}\mov(0,#2.1){\hr*{.988}}\relax
\mov(0,#2.15){\hr*{.969}}\mov(0,#2.2){\hr*{.937}}\relax
\mov(0,#2.25){\hr*{.893}}\mov(0,#2.3){\hr*{.836}}\relax
\mov(0,#2.35){\hr*{.760}}\mov(0,#2.4){\hr*{.662}}\relax
\mov(0,#2.45){\hr*{.531}}\mov(0,#2.5){\hr*{.320}}\relax
\special{em:linewidth \the\linwid*}}}}

\def\vdot{\rlap{\kern-1.9pt\lower1.8pt\hbox{$\scriptstyle\bullet$}}}
\def\vtimes{\rlap{\kern-3pt\lower1.8pt\hbox{$\scriptstyle\times$}}}
\def\vDot{\rlap{\kern-2.3pt\lower2.7pt\hbox{$\bullet$}}}
\def\vTimes{\rlap{\kern-3.6pt\lower2.4pt\hbox{$\times$}}}

\def\arc(#1)[#2,#3]{{\k*=#2\l*=#3\m*=\l*
\advance\m*-6\ifnum\k*>\l*\relax\else
{\rotate(#2)\mov(#1,0){\sm*}}\loop
\ifnum\k*<\m*\advance\k*5{\rotate(\k*)\mov(#1,0){\sl*}}\repeat
{\rotate(#3)\mov(#1,0){\sl*}}\fi}}

\def\dasharc(#1)[#2,#3]{{\k**=#2\n*=#3\advance\n*-1\advance\n*-\k**
\L*=1000sp\L*#1\L* \multiply\L*\n* \multiply\L*\Nhalfperiods
\divide\L*57\N*\L* \divide\N*2000\ifnum\N*=0\N*1\fi
\r*\n*  \divide\r*\N* \ifnum\r*<2\r*2\fi
\m**\r* \divide\m**2 \l**\r* \advance\l**-\m** \N*\n* \divide\N*\r*
\k**\r* \multiply\k**\N* \dn*\n* \advance\dn*-\k** \divide\dn*2\advance\dn*\*one
\r*\l** \divide\r*2\advance\dn*\r* \advance\N*-2\k**#2\relax
\ifnum\l**<6{\rotate(#2)\mov(#1,0){\sm*}}\advance\k**\dn*
{\rotate(\k**)\mov(#1,0){\sl*}}\advance\k**\m**
{\rotate(\k**)\mov(#1,0){\sm*}}\loop
\advance\k**\l**{\rotate(\k**)\mov(#1,0){\sl*}}\advance\k**\m**
{\rotate(\k**)\mov(#1,0){\sm*}}\advance\N*-1\ifnum\N*>0\repeat
{\rotate(#3)\mov(#1,0){\sl*}}\else\advance\k**\dn*
\arc(#1)[#2,\k**]\loop\advance\k**\m** \r*\k**
\advance\k**\l** {\arc(#1)[\r*,\k**]}\relax
\advance\N*-1\ifnum\N*>0\repeat
\advance\k**\m**\arc(#1)[\k**,#3]\fi}}

\def\triangarc#1(#2)[#3,#4]{{\k**=#3\n*=#4\advance\n*-\k**
\L*=1000sp\L*#2\L* \multiply\L*\n* \multiply\L*\Nhalfperiods
\divide\L*57\N*\L* \divide\N*1000\ifnum\N*=0\N*1\fi
\d**=#2\Lengthunit \d*\d** \divide\d*57\multiply\d*\n*
\r*\n*  \divide\r*\N* \ifnum\r*<2\r*2\fi
\m**\r* \divide\m**2 \l**\r* \advance\l**-\m** \N*\n* \divide\N*\r*
\dt*\d* \divide\dt*\N* \dt*.5\dt* \dt*#1\dt*
\divide\dt*1000\multiply\dt*\magnitude
\k**\r* \multiply\k**\N* \dn*\n* \advance\dn*-\k** \divide\dn*2\relax
\r*\l** \divide\r*2\advance\dn*\r* \advance\N*-1\k**#3\relax
{\rotate(#3)\mov(#2,0){\sm*}}\advance\k**\dn*
{\rotate(\k**)\mov(#2,0){\sl*}}\advance\k**-\m**\advance\l**\m**\loop\dt*-\dt*
\d*\d** \advance\d*\dt*
\advance\k**\l**{\rotate(\k**)\rmov*(\d*,0pt){\sl*}}%
\advance\N*-1\ifnum\N*>0\repeat\advance\k**\m**
{\rotate(\k**)\mov(#2,0){\sl*}}{\rotate(#4)\mov(#2,0){\sl*}}}}

\def\wavearc#1(#2)[#3,#4]{{\k**=#3\n*=#4\advance\n*-\k**
\L*=4000sp\L*#2\L* \multiply\L*\n* \multiply\L*\Nhalfperiods
\divide\L*57\N*\L* \divide\N*1000\ifnum\N*=0\N*1\fi
\d**=#2\Lengthunit \d*\d** \divide\d*57\multiply\d*\n*
\r*\n*  \divide\r*\N* \ifnum\r*=0\r*1\fi
\m**\r* \divide\m**2 \l**\r* \advance\l**-\m** \N*\n* \divide\N*\r*
\dt*\d* \divide\dt*\N* \dt*.7\dt* \dt*#1\dt*
\divide\dt*1000\multiply\dt*\magnitude
\k**\r* \multiply\k**\N* \dn*\n* \advance\dn*-\k** \divide\dn*2\relax
\divide\N*4\advance\N*-1\k**#3\relax
{\rotate(#3)\mov(#2,0){\sm*}}\advance\k**\dn*
{\rotate(\k**)\mov(#2,0){\sl*}}\advance\k**-\m**\advance\l**\m**\loop\dt*-\dt*
\d*\d** \advance\d*\dt* \dd*\d** \advance\dd*1.3\dt*
\advance\k**\r*{\rotate(\k**)\rmov*(\d*,0pt){\sl*}}\relax
\advance\k**\r*{\rotate(\k**)\rmov*(\dd*,0pt){\sl*}}\relax
\advance\k**\r*{\rotate(\k**)\rmov*(\d*,0pt){\sl*}}\relax
\advance\k**\r*
\advance\N*-1\ifnum\N*>0\repeat\advance\k**\m**
{\rotate(\k**)\mov(#2,0){\sl*}}{\rotate(#4)\mov(#2,0){\sl*}}}}

\def\gmov*#1(#2,#3)#4{\rlap{\L*=#1\Lengthunit
\xL*=#2\L* \yL*=#3\L*
\rx* \gcos*\xL* \tmp* \gsin*\yL* \advance\rx*-\tmp*
\ry* \gcos*\yL* \tmp* \gsin*\xL* \advance\ry*\tmp*
\rx*=\xscale\rx* \ry*=\yscale\ry*
\xL* \the\cos*\rx* \tmp* \the\sin*\ry* \advance\xL*-\tmp*
\yL* \the\cos*\ry* \tmp* \the\sin*\rx* \advance\yL*\tmp*
\kern\xL*\raise\yL*\hbox{#4}}}

\def\rgmov*(#1,#2)#3{\rlap{\xL*#1\yL*#2\relax
\rx* \gcos*\xL* \tmp* \gsin*\yL* \advance\rx*-\tmp*
\ry* \gcos*\yL* \tmp* \gsin*\xL* \advance\ry*\tmp*
\rx*=\xscale\rx* \ry*=\yscale\ry*
\xL* \the\cos*\rx* \tmp* \the\sin*\ry* \advance\xL*-\tmp*
\yL* \the\cos*\ry* \tmp* \the\sin*\rx* \advance\yL*\tmp*
\kern\xL*\raise\yL*\hbox{#3}}}

\def\Earc(#1)[#2,#3][#4,#5]{{\k*=#2\l*=#3\m*=\l*
\advance\m*-6\ifnum\k*>\l*\relax\else\def\xscale{#4}\def\yscale{#5}\relax
{\angle**0\rotate(#2)}\gmov*(#1,0){\sm*}\loop
\ifnum\k*<\m*\advance\k*5\relax
{\angle**0\rotate(\k*)}\gmov*(#1,0){\sl*}\repeat
{\angle**0\rotate(#3)}\gmov*(#1,0){\sl*}\relax
\def\xscale{1}\def\yscale{1}\fi}}

\def\dashEarc(#1)[#2,#3][#4,#5]{{\k**=#2\n*=#3\advance\n*-1\advance\n*-\k**
\L*=1000sp\L*#1\L* \multiply\L*\n* \multiply\L*\Nhalfperiods
\divide\L*57\N*\L* \divide\N*2000\ifnum\N*=0\N*1\fi
\r*\n*  \divide\r*\N* \ifnum\r*<2\r*2\fi
\m**\r* \divide\m**2 \l**\r* \advance\l**-\m** \N*\n* \divide\N*\r*
\k**\r*\multiply\k**\N* \dn*\n* \advance\dn*-\k** \divide\dn*2\advance\dn*\*one
\r*\l** \divide\r*2\advance\dn*\r* \advance\N*-2\k**#2\relax
\ifnum\l**<6\def\xscale{#4}\def\yscale{#5}\relax
{\angle**0\rotate(#2)}\gmov*(#1,0){\sm*}\advance\k**\dn*
{\angle**0\rotate(\k**)}\gmov*(#1,0){\sl*}\advance\k**\m**
{\angle**0\rotate(\k**)}\gmov*(#1,0){\sm*}\loop
\advance\k**\l**{\angle**0\rotate(\k**)}\gmov*(#1,0){\sl*}\advance\k**\m**
{\angle**0\rotate(\k**)}\gmov*(#1,0){\sm*}\advance\N*-1\ifnum\N*>0\repeat
{\angle**0\rotate(#3)}\gmov*(#1,0){\sl*}\def\xscale{1}\def\yscale{1}\else
\advance\k**\dn* \Earc(#1)[#2,\k**][#4,#5]\loop\advance\k**\m** \r*\k**
\advance\k**\l** {\Earc(#1)[\r*,\k**][#4,#5]}\relax
\advance\N*-1\ifnum\N*>0\repeat
\advance\k**\m**\Earc(#1)[\k**,#3][#4,#5]\fi}}

\def\triangEarc#1(#2)[#3,#4][#5,#6]{{\k**=#3\n*=#4\advance\n*-\k**
\L*=1000sp\L*#2\L* \multiply\L*\n* \multiply\L*\Nhalfperiods
\divide\L*57\N*\L* \divide\N*1000\ifnum\N*=0\N*1\fi
\d**=#2\Lengthunit \d*\d** \divide\d*57\multiply\d*\n*
\r*\n*  \divide\r*\N* \ifnum\r*<2\r*2\fi
\m**\r* \divide\m**2 \l**\r* \advance\l**-\m** \N*\n* \divide\N*\r*
\dt*\d* \divide\dt*\N* \dt*.5\dt* \dt*#1\dt*
\divide\dt*1000\multiply\dt*\magnitude
\k**\r* \multiply\k**\N* \dn*\n* \advance\dn*-\k** \divide\dn*2\relax
\r*\l** \divide\r*2\advance\dn*\r* \advance\N*-1\k**#3\relax
\def\xscale{#5}\def\yscale{#6}\relax
{\angle**0\rotate(#3)}\gmov*(#2,0){\sm*}\advance\k**\dn*
{\angle**0\rotate(\k**)}\gmov*(#2,0){\sl*}\advance\k**-\m**
\advance\l**\m**\loop\dt*-\dt* \d*\d** \advance\d*\dt*
\advance\k**\l**{\angle**0\rotate(\k**)}\rgmov*(\d*,0pt){\sl*}\relax
\advance\N*-1\ifnum\N*>0\repeat\advance\k**\m**
{\angle**0\rotate(\k**)}\gmov*(#2,0){\sl*}\relax
{\angle**0\rotate(#4)}\gmov*(#2,0){\sl*}\def\xscale{1}\def\yscale{1}}}

\def\waveEarc#1(#2)[#3,#4][#5,#6]{{\k**=#3\n*=#4\advance\n*-\k**
\L*=4000sp\L*#2\L* \multiply\L*\n* \multiply\L*\Nhalfperiods
\divide\L*57\N*\L* \divide\N*1000\ifnum\N*=0\N*1\fi
\d**=#2\Lengthunit \d*\d** \divide\d*57\multiply\d*\n*
\r*\n*  \divide\r*\N* \ifnum\r*=0\r*1\fi
\m**\r* \divide\m**2 \l**\r* \advance\l**-\m** \N*\n* \divide\N*\r*
\dt*\d* \divide\dt*\N* \dt*.7\dt* \dt*#1\dt*
\divide\dt*1000\multiply\dt*\magnitude
\k**\r* \multiply\k**\N* \dn*\n* \advance\dn*-\k** \divide\dn*2\relax
\divide\N*4\advance\N*-1\k**#3\def\xscale{#5}\def\yscale{#6}\relax
{\angle**0\rotate(#3)}\gmov*(#2,0){\sm*}\advance\k**\dn*
{\angle**0\rotate(\k**)}\gmov*(#2,0){\sl*}\advance\k**-\m**
\advance\l**\m**\loop\dt*-\dt*
\d*\d** \advance\d*\dt* \dd*\d** \advance\dd*1.3\dt*
\advance\k**\r*{\angle**0\rotate(\k**)}\rgmov*(\d*,0pt){\sl*}\relax
\advance\k**\r*{\angle**0\rotate(\k**)}\rgmov*(\dd*,0pt){\sl*}\relax
\advance\k**\r*{\angle**0\rotate(\k**)}\rgmov*(\d*,0pt){\sl*}\relax
\advance\k**\r*
\advance\N*-1\ifnum\N*>0\repeat\advance\k**\m**
{\angle**0\rotate(\k**)}\gmov*(#2,0){\sl*}\relax
{\angle**0\rotate(#4)}\gmov*(#2,0){\sl*}\def\xscale{1}\def\yscale{1}}}

\newcount\CatcodeOfAtSign
\CatcodeOfAtSign=\the\catcode`\@
\catcode`\@=11
\def\@arc#1[#2][#3]{\rlap{\Lengthunit=#1\Lengthunit
\sm*\l*arc(#2.1914,#3.0381)[#2][#3]\relax
\mov(#2.1914,#3.0381){\l*arc(#2.1622,#3.1084)[#2][#3]}\relax
\mov(#2.3536,#3.1465){\l*arc(#2.1084,#3.1622)[#2][#3]}\relax
\mov(#2.4619,#3.3086){\l*arc(#2.0381,#3.1914)[#2][#3]}}}

\def\dash@arc#1[#2][#3]{\rlap{\Lengthunit=#1\Lengthunit
\d*arc(#2.1914,#3.0381)[#2][#3]\relax
\mov(#2.1914,#3.0381){\d*arc(#2.1622,#3.1084)[#2][#3]}\relax
\mov(#2.3536,#3.1465){\d*arc(#2.1084,#3.1622)[#2][#3]}\relax
\mov(#2.4619,#3.3086){\d*arc(#2.0381,#3.1914)[#2][#3]}}}

\def\wave@arc#1[#2][#3]{\rlap{\Lengthunit=#1\Lengthunit
\w*lin(#2.1914,#3.0381)\relax
\mov(#2.1914,#3.0381){\w*lin(#2.1622,#3.1084)}\relax
\mov(#2.3536,#3.1465){\w*lin(#2.1084,#3.1622)}\relax
\mov(#2.4619,#3.3086){\w*lin(#2.0381,#3.1914)}}}

\def\bezier#1(#2,#3)(#4,#5)(#6,#7){\N*#1\l*\N* \advance\l*\*one
\d* #4\Lengthunit \advance\d* -#2\Lengthunit \multiply\d* \*two
\b* #6\Lengthunit \advance\b* -#2\Lengthunit
\advance\b*-\d* \divide\b*\N*
\d** #5\Lengthunit \advance\d** -#3\Lengthunit \multiply\d** \*two
\b** #7\Lengthunit \advance\b** -#3\Lengthunit
\advance\b** -\d** \divide\b**\N*
\mov(#2,#3){\sm*{\loop\ifnum\m*<\l*
\a*\m*\b* \advance\a*\d* \divide\a*\N* \multiply\a*\m*
\a**\m*\b** \advance\a**\d** \divide\a**\N* \multiply\a**\m*
\rmov*(\a*,\a**){\unhcopy\spl*}\advance\m*\*one\repeat}}}

\catcode`\*=12

\newcount\n@ast
\def\n@ast@#1{\n@ast0\relax\get@ast@#1\end}
\def\get@ast@#1{\ifx#1\end\let\next\relax\else
\ifx#1*\advance\n@ast1\fi\let\next\get@ast@\fi\next}

\newif\if@up \newif\if@dwn
\def\up@down@#1{\@upfalse\@dwnfalse
\if#1u\@uptrue\fi\if#1U\@uptrue\fi\if#1+\@uptrue\fi
\if#1d\@dwntrue\fi\if#1D\@dwntrue\fi\if#1-\@dwntrue\fi}

\def\halfcirc#1(#2)[#3]{{\Lengthunit=#2\Lengthunit\up@down@{#3}\relax
\if@up\mov(0,.5){\@arc[-][-]\@arc[+][-]}\fi
\if@dwn\mov(0,-.5){\@arc[-][+]\@arc[+][+]}\fi
\def\lft{\mov(0,.5){\@arc[-][-]}\mov(0,-.5){\@arc[-][+]}}\relax
\def\rght{\mov(0,.5){\@arc[+][-]}\mov(0,-.5){\@arc[+][+]}}\relax
\if#3l\lft\fi\if#3L\lft\fi\if#3r\rght\fi\if#3R\rght\fi
\n@ast@{#1}\relax
\ifnum\n@ast>0\if@up\shade[+]\fi\if@dwn\shade[-]\fi\fi
\ifnum\n@ast>1\if@up\dshade[+]\fi\if@dwn\dshade[-]\fi\fi}}

\def\halfdashcirc(#1)[#2]{{\Lengthunit=#1\Lengthunit\up@down@{#2}\relax
\if@up\mov(0,.5){\dash@arc[-][-]\dash@arc[+][-]}\fi
\if@dwn\mov(0,-.5){\dash@arc[-][+]\dash@arc[+][+]}\fi
\def\lft{\mov(0,.5){\dash@arc[-][-]}\mov(0,-.5){\dash@arc[-][+]}}\relax
\def\rght{\mov(0,.5){\dash@arc[+][-]}\mov(0,-.5){\dash@arc[+][+]}}\relax
\if#2l\lft\fi\if#2L\lft\fi\if#2r\rght\fi\if#2R\rght\fi}}

\def\halfwavecirc(#1)[#2]{{\Lengthunit=#1\Lengthunit\up@down@{#2}\relax
\if@up\mov(0,.5){\wave@arc[-][-]\wave@arc[+][-]}\fi
\if@dwn\mov(0,-.5){\wave@arc[-][+]\wave@arc[+][+]}\fi
\def\lft{\mov(0,.5){\wave@arc[-][-]}\mov(0,-.5){\wave@arc[-][+]}}\relax
\def\rght{\mov(0,.5){\wave@arc[+][-]}\mov(0,-.5){\wave@arc[+][+]}}\relax
\if#2l\lft\fi\if#2L\lft\fi\if#2r\rght\fi\if#2R\rght\fi}}

\catcode`\*=11

\def\Circle#1(#2){\halfcirc#1(#2)[u]\halfcirc#1(#2)[d]\n@ast@{#1}\relax
\ifnum\n@ast>0\L*=\xscale\Lengthunit
\ifnum\angle**=0\clap{\vrule width#2\L* height.1pt}\else
\L*=#2\L*\L*=.5\L*\special{em:linewidth .001pt}\relax
\rmov*(-\L*,0pt){\sm*}\rmov*(\L*,0pt){\sl*}\relax
\special{em:linewidth \the\linwid*}\fi\fi}

\catcode`\*=12

\def\wavecirc(#1){\halfwavecirc(#1)[u]\halfwavecirc(#1)[d]}

\def\dashcirc(#1){\halfdashcirc(#1)[u]\halfdashcirc(#1)[d]}

\def\xscale{1}
\def\yscale{1}

\def\Ellipse#1(#2)[#3,#4]{\def\xscale{#3}\def\yscale{#4}\relax
\Circle#1(#2)\def\xscale{1}\def\yscale{1}}

\def\dashEllipse(#1)[#2,#3]{\def\xscale{#2}\def\yscale{#3}\relax
\dashcirc(#1)\def\xscale{1}\def\yscale{1}}

\def\waveEllipse(#1)[#2,#3]{\def\xscale{#2}\def\yscale{#3}\relax
\wavecirc(#1)\def\xscale{1}\def\yscale{1}}

\def\halfEllipse#1(#2)[#3][#4,#5]{\def\xscale{#4}\def\yscale{#5}\relax
\halfcirc#1(#2)[#3]\def\xscale{1}\def\yscale{1}}

\def\halfdashEllipse(#1)[#2][#3,#4]{\def\xscale{#3}\def\yscale{#4}\relax
\halfdashcirc(#1)[#2]\def\xscale{1}\def\yscale{1}}

\def\halfwaveEllipse(#1)[#2][#3,#4]{\def\xscale{#3}\def\yscale{#4}\relax
\halfwavecirc(#1)[#2]\def\xscale{1}\def\yscale{1}}

\catcode`\@=\the\CatcodeOfAtSign
\thispagestyle{empty}
\renewcommand{\thefootnote}{\alph{footnote}}

\begin{flushright}
hep-th/0210241
\vspace{1.5cm}\\ 
\end{flushright}

\begin{center}
{\Large\bf  Complete Low-Energy Effective
Action in ${\cal N}$=4 SYM:}
\vspace{0.1cm}

{\Large\bf a Direct ${\cal N}=2$ Supergraph Calculation}
\end{center}
\begin{center}
{\bf I.L. Buchbinder $^+$\footnote{$\,$joseph@tspu.edu.ru}, 
E.A. Ivanov $^*$\footnote{$\,$eivanov@thsun1.jinr.ru}, 
A.Yu. Petrov 
$^{+\dagger}$\footnote{$\,$petrov@fma.if.usp.br, petrov@tspu.edu.ru} }

\vspace{2mm}

${}^+$\footnotesize{ {\it Department of Theoretical Physics\\
Tomsk State Pedagogical University\\
Tomsk 634041, Russia}}

\vspace{.2cm}

${}^*$ \footnotesize { {\it Bogoliubov Laboratory of Theoretical Physics,\\
Joint Institute for Nuclear Research,\\
Dubna, 141980 Moscow Region, Russia}}\\

\vspace{0.2cm}

${}^\dagger$\footnotesize{
{\it
Instituto de Fisica, Universidade de S\~{a}o Paulo\\
P.O. Box 66318, 05315-970, S\~{a}o Paulo, Brasil}}
\end{center}

\begin{abstract}
Using the covariant $\N=2$ harmonic supergraph techniques we calculate the one-loop low-energy
effective action  of $\N=4$ super-Yang-Mills theory in Coulomb branch with gauge group $SU(2)$
spontaneously broken down to $U(1)$. The full dependence of the low-energy effective action on both
the hypermultiplet and gauge fields is determined. The direct quantum calculation confirms
the correctness of the exact $\N=4$ SYM low-energy effective action derived in hep-th/0111062 on the purely
algebraic ground by invoking a hidden $\N=2$ supersymmetry which completes the manifest
$\N=2$ one to $\N=4$. Our results provide an exhaustive solution to the problem of finding out
the exact completely $\N=4$ supersymmetric low-energy effective action for the theory under consideration.
\end{abstract}

\newpage
\setcounter{page}{1}
\setcounter{footnote}{0}
\renewcommand{\thefootnote}{\arabic{footnote}}
\section{Introduction}
One of the remarkable features of ${\cal N}=4$ supersymmetric
quantum Yang-Mills theory is the opportunity to obtain exact
results. At present, one can distinguish at least three trends in
finding out exact solutions to some important quantities in this theory.
These are, first, the study of low-energy effective action, second,
computing the correlators of gauge invariant operators, and, third,
computing the expectation values of Wilson loops (see \cite{rev}-\cite{loop}
for a review and references).

In this paper we solve the open problem of calculating the exact
low-energy effective action depending on all fields of
${\cal N}=4$ gauge multiplet. Until recently, the well established exact results
for leading low-energy contributions to the effective action
were obtained for $SU(2)$, ${\cal N}=4$ SYM model in Coulomb branch
\cite{dine,bbk,bbkl} and only for the ${\cal N}=2$ gauge
multiplet sector of the effective action. These contributions are presented by
a non-holomorphic effective potential of the form
\footnote {A possibility of non-holomorphic contributions having the form of product of two
logarithms was earlier found in \cite{rocek} for the effective action
of a generic ${\cal N}=2$ SYM model.}
\bea
{\cal H}(W, \bar{W}) = \frac{1}{(4\pi)^2} \ln\frac{W}{\L}
\ln\frac{\bar{W}}{\L}~.  \label{nonhol}
\eea
Here $W$ and $\bar{W}$ are the ${\cal N}=2$ $U(1)$ gauge superfield strengths and
$\Lambda$ is an arbitrary scale. It was pointed out in \cite{dine,dg}
that although the potential (\ref{nonhol}) was obtained in one-loop
approximation, it receives neither
perturbative nor non-perturbative corrections. Hence, the function
${\cal H}(W,\bar{W})$ (\ref{nonhol}) determines the exact low-energy
effective action in the ${\cal N}=2$ gauge field sector (to be more precise,
the leading in external momenta part of the full effective action).

A generalization of the effective potential (\ref{nonhol})
to the Coulomb branch of $\N=4$ SYM model with the gauge group $SU(N)$
broken to its Cartan subgroup $U(1)^{N-1}$ has been given in
\cite{dg,bbk2,bbka}
(see also the review \cite{rev}). Despite the fact that the one-loop non-holomorphic potential
in this case looks quite analogous to (\ref{nonhol}), it was argued in
\cite{dg,dg1} that the effective potential can contain, in principle, some extra
non-logarithmic contributions (begining at least with fifth loop \cite{bp})
which do not have the product form (\ref{nonhol}). However, in the actual computations
such contributions have never  been found.

From the standpoint of ${\cal N}=2$ supersymmetry, the ${\cal N}=4$ gauge multiplet is
a sum of ${\cal N}=2$ gauge multiplet and hypermultiplet. All the
above mentioned results on the structure of non-holomorphic
potential were obtained only for that part of the effective action which depends
on the fields of ${\cal N}=2$ gauge multiplet. The problem of constructing
the {\it complete} effective action depending on both the ${\cal N}=2$ gauge
multiplet and hypermultiplet fields has been solved in a recent paper
\cite{BI}. The construction of \cite{BI} is based upon the purely algebraic analysis exploring the
existence of extra hidden on-shell ${\cal N}=2$ supersymmetry in $\N=4$ SYM theory
formulated in $\N=2$ harmonic superspace \cite{GIKOS,GIKOS3}. The manifest
off-shell ${\cal N}=2$ supersymmetry and the hidden one constitute
the full on-shell ${\cal N}=4$ supersymmetry of ${\cal N}=4$
SYM theory. It was shown that the
potential (\ref{nonhol}), as well as its generalization to $SU(N)$ model,
can be completed to ${\cal N}=4$ supersymmetric form by
adding the appropriate terms depending on hypermultiplet superfields. However,
any non-logarithmic terms in the low-energy effective action do not admit such a completion
and hence are ruled out by ${\cal N}=4$ supersymmetry. An open problem
to remain was to re-derive the effective action of \cite{BI} by a direct supergraph
computation in the quantum field theory framework. The present paper is devoted to solving this
problem.

We specialize to the $SU(2)$ ${\cal N}=4$ SYM model formulated in
harmonic superspace and consider its Coulomb branch. Since the effective potential
(\ref{nonhol}) is generated solely by one-loop contributions, the corresponding
hypermultiplet-dependent terms
also have to be evaluated only in the one-loop approximation. We would
like to point out that such a calculation is a very non-trivial
problem. We have to calculate not just a contribution of single supergraph with a
few external legs but the contributions of the supergraphs with
arbitrary numbers of external legs on a background of non-zero $W$ and
$\bar W$ and then to sum up all these contributions. We show that such calculations
can be actually accomplished and the result yields
just the effective action found in \cite{BI}. A similar approach
is used to give a new derivation of the one-loop
non-holomorphic effective $W, \bar W$ potential. As the result
we are aware of the unified approach allowing us to compute the low-energy effective action
in both the gauge multiplet and hypermultiplet sectors. The calculations are
carried out within the ${\cal N}=2$ background field method \cite{bko},
with making use of the quantum ${\cal N}=2$ harmonic superfield techniques
pioneered in \cite{GIKOS} and further advanced in \cite{bp},
\cite{bbiko}--\cite{kuz1}.

The paper is organized as follows. In section 2 we recapitulate
a formulation of $\N=4$ SYM theory in $\N =2$ harmonic superspace
and describe a general structure of low-energy effective action in this
theory. Section 3 presents details of calculating
the hypermultiplet-dependent contributions to effective action.
In section 4, using a similar approach, we perform a manifestly
$\N=2$ supercovariant calculation of the non-holomorphic effective
potential in the $\N=2$ gauge fields sector. The results are summarized
in section 5.
\setcounter{equation}{0}

\section{${\cal N}=4$ SYM theory in harmonic superspace and
the problem of complete low-energy effective action}
The `microscopic' action of ${\cal
N}=4$ $SU(2)$ SYM theory written in ${\cal N}=2$ harmonic superspace
\cite{GIKOS,GIKOS3} reads
\bea \label{2} S= \frac{1}{2g^2}{\rm tr}\int d^8 z\,
{\cal W}^2-\frac{1}{2}{\rm tr}\int d\zeta^{(-4)} q^{+a}(D^{++}+iV^{++})q^+_a\,.
\eea
Here $q^+_a=(q^+,\tilde{q}^+)$, $q^{+a}=\epsilon^{ab}q^+_b$, is the
hypermultiplet in the adjoint representation of the gauge group and
${\cal W}$ is
the covariant strength of ${\cal N}=2$ analytic gauge superfield $V^{++}$.
This action (i) is manifestly $\N=2$ supersymmetric; (ii) possesses a second hidden $\N=2$
supersymmetry completing the first one to ${\cal N}=4$. The off-shell
$\N=4$ transformations for the full nonabelian case can be found in
\cite{GIKOS,GIKOS3}. \footnote{To avoid a possible confusion we note that,
though these hidden supersymmetry transformations can be defined for
off-shell superfields, they form, together with the manifest $\N=2$
supersymmetry transformations, $\N=4$ Poincar\'e supersymmetry only on shell.}
For fixing the one-loop effective action it suffices to know them
for the abelian $U(1)$ components of ${\cal W}, q^{+a}$ only 
(with all off-diagonal
superfields neglected) and on the free mass shell for these $U(1)$
superfields \cite{BI}:
\bea \label{hidden}
\delta {\cal W} &=&
\frac{1}{2}\bar{\e}^{\ad a} \bar{D}^-_{\ad}q^+_a\,,\quad \delta
\bar{\cal W} =
\frac{1}{2}\e^{\a a} D^-_{\a}q^+_a\,, \nonumber\\
\delta
q^{+}_{a}&=&\frac{1}{4}(\e^{\b}_a D^+_{\b}{\cal W} +
\bar{\e}^{\ad}_a\bar{D}^+_{\ad}\bar{\cal W})\,, \quad \delta
q^{-}_a=\frac{1}{4}(\e^{\b}_a D^-_{\b}{\cal W} +
\bar{\e}^{\ad}_a\bar{D}^-_{\ad}\bar{\cal W})\,.
\eea
Here $q^{- a} \equiv D^{--}q^{+ a}\,$, where $D^{--}$ is the
second harmonic derivative.

For calculating quantum corrections we use the background field
method \cite{bko}. We start with carrying
out background-quantum splitting by the rule
$q^{+a}\to q^{+a}+Q^{+a};\, V^{++}\to V^{++}+gv^{++}$.
Here $q^{+a}$ is a background hypermultiplet and
$Q^{+a}$ a quantum one; $V^{++}$ is a background $\N=2$
gauge superfield and $v^{++}$ a quantum one.
For our purpose it suffices to retain in the action
(\ref{2}) only terms of the second order in quantum superfields:
\bea
S_2&=&
-\frac{1}{2}{\rm tr}\int d\zeta^{(-4)}
\Big[v^{++}\stackrel{\frown}{\Box}v^{++}+
Q^{+a}(D^{++}+iV^{++})Q^+_a+\nonumber\\&+&
Q^{+a}(igv^{++})q^+_a
+q^{+a}(igv^{++})Q^+_a
\Big]\,. \label{S21}
\eea
The operator $\stackrel{\frown}{\Box}$ includes background ${\cal N}=2$
superfield strengths ${\cal W},\bar{\cal W}$, is defined to act in the adjoint representation
of the gauge group and has the form \cite{bbiko,bt2}:
\bea
\label{box0}
\stackrel{\frown}{\Box}&=&\Box+\frac{i}{2}({\cal D}^{+\a}{\cal W}){\cal
D}^-_{\a}+\frac{i}{2}(\bar{\cal D}^+_{\ad}\bar{\cal W})\bar{\cal D}^{-\ad}+
\frac{i}{8}[{\cal D}^{+\a},{\cal D}^-_{\a}]{\cal W} \nonumber\\&-&
\frac{i}{4}({\cal D}^{+\a}{\cal D}^+_{\a}) {\cal W} D^{--}
+\frac{1}{2}\{{\cal W},\bar{\cal W}\}~.
\eea

The $\N=4$ SYM effective action in the Coulomb branch depends only on
abelian background superfields ${\cal W},\bar{\cal W}\,$, 
$q^{+a}$ which can be chosen
to be on-shell, $[{\cal D}^{+\a},{\cal D}^-_{\a}]{\cal W}=0\,$, \\
$({\cal D}^{+\a}{\cal D}^+_{\a}) {\cal W} =0\,$, $D^{++}q^{+a} =0\,$ 
\cite{bt2,BI}.
Taking into account these conditions and the fact that all superfields are
in the adjoint representation, we have
\bea
\label{ad}
\stackrel{\frown}{\Box} v^{++}&=&
\Box v^{++}+\frac{i}{2}[{\cal D}^{+\a} {\cal W},{\cal D}^-_{\a}v^{++}]+
\frac{i}{2}[\bar{\cal D}^{+\ad}\bar{\cal W},\bar{\cal D}^-_{\ad}v^{++}]+
[{\cal W},[\bar{\cal W},v^{++}]]~,\nonumber\\
Q^{+a}v^{++}Q^+_{a}&=&
Q^{+a}[v^{++},Q^+_{a}]~, \ \mbox{etc}~.
\eea
The quantum fields $v^{++}$ with values in $su(2)$
can be written as
\bea
\label{exgen10}
v^{++} =v^{++}_{i}\,\t_i~,
\eea
where  $\t_i$
are generators of $su(2)$ related to Pauli matrices $\s_i$ as
\bea
\label{gen10}
\t_i=\frac{1}{\sqrt{2}}\s_i~.
\eea
They satisfy the relations
\bea
\label{rela}
[\t_i,\t_j]=i\sqrt{2}\e_{ijk}\t_k~,\quad {\rm tr}(\t_i\t_j)=\delta_{ij}~.
\eea

The procedure of calculation of quantum correction for the gauge group
$SU(N)$ broken to $U(1)$ was developed in \cite{bt2}.
The background $\N=2$ strengths are expanded as
${\cal W}={\cal W}_i\t_i$.
We calculate the effective action in the Coulomb branch, in which it
depends only on the  background fields taking values
in the unbroken $u(1)$. The latter corresponds to the restriction
${\cal W}={\cal W}_3\t_3$, further we denote ${\cal W}_3$ as $W$.
Our first aim will be to derive the action (and hence, vertices and
propagators) for fields $v^{++}_i,\,Q^{+a}_i$.
We also redefine $gq^{+a}\to q^{+a}$. For the Abelian background we have
${\cal D}^{+\a} W=D^{+\a} W\,$, $\bar{\cal D}^{+\ad}
\bar{W}=\bar{D}^{+\ad}\bar{W}\,$, with $D,\bar{D}$ being flat
spinor derivatives, and we can make these replacements in (\ref{box0}).

Thus the quadratic part of the action can be expressed as
\bea
\label{qu2}
S_2
&=&-\frac{1}{2}
\int d\zeta^{(-4)}\Big\{v^{++}_i\Box v^{++}_j\ {\rm tr}(\t_i\t_j)+
Q^{+a}_iD^{++}Q^+_{aj}\ {\rm tr}(\t_i\t_j) \nonumber\\
&& +\,
Q^{+a}_i(iV^{++})Q^+_{aj}{\rm tr}(\t_i[\t_3,\t_j])+
v^{++}_iW\bar{W} v^{++}_j\
{\rm tr}[\t_i[\t_3,[\t_3,\t_j]]
\nonumber\\
& &+\frac{i}{2}v^{++}_i(D^{+\a}W){\cal D}^-_{\a}v^{++}_j
{\rm tr}(\t_i[\t_3,\t_j])+
\frac{i}{2}v^{++}_i(\bar{D}^{+\ad}W)\bar{\cal D}^-_{\ad}v^{++}_j
{\rm tr}(\t_i[\t_3,\t_j])\nonumber\\
&&+\,
iv^{++}_i\Big(q^{+a}Q^+_{aj}{\rm tr}(\t_3[\t_i,\t_j])+
Q^+_{aj}q^{+a}{\rm tr}(\t_i[\t_j,\t_3])\ \Big)
\Big\}\,.
\eea
Using (\ref{gen10}), we can do all
traces. After this the expression (\ref{qu2}) is rewritten as
\bea
\label{sq}
S_2
&=&-\hf
\int d\zeta^{(-4)} \Big\{
 v^{++}_1(\Box +2W\bar{W})v^{++}_1+
v^{++}_2(\Box +2W\bar{W})v^{++}_2-
\nonumber\\
&-&\frac{1}{\sqrt{2}}v^{++}_1((D^{+\a}W){\cal D}^-_{\a}+
(\bar{D}^{+\ad}W)\bar{\cal D}^-_{\ad})v^{++}_2+
\frac{1}{\sqrt{2}}v^{++}_2((D^{+\a}W){\cal D}^-_{\a}+\nonumber\\&+&
(\bar{D}^{+\ad}W)\bar{\cal D}^-_{\ad})v^{++}_1
+v^{++}_3\Box v^{++}_3+Q^{+i}_3D^{++}Q^+_{i3}+
\nonumber\\&&+\,
Q^{+a}_1(D^{++}+i\sqrt{2}V^{++})Q^+_{a2}-
Q^{+a}_2(D^{++}+i\sqrt{2}V^{++})Q^+_{a1}+
\nonumber\\&& + \,
iv^{++}_2\sqrt{2}(Q^{+a}_1q^+_a+q^+_aQ^{+a}_1)-
iv^{++}_1\sqrt{2}(Q^{+a}_2 q^+_a+q^+_aQ^{+a}_2)
\Big\}\,.
\eea
 From (\ref{sq}) we  conclude that only the components carrying indices 1,2
have a non-trivial background-dependent propagator,
while the quantum superfield components with index 3
do not interact with the background, totally decouple from the action
and possess the free propagators.
Then we can define new complex quantum fields
\bea
& &\c^{++}=\frac{1}{\sqrt{2}}(v^{++}_1-iv^{++}_2)~,\;\;
\bar{\c}^{++}=\frac{1}{\sqrt{2}}(v^{++}_1+iv^{++}_2)~,\nonumber\\
& &\eta^{+i}=\frac{1}{\sqrt{2}}(Q^{+i}_1+iQ^{+i}_2)~,\;\;
\bar{\eta}^{+i}=\frac{1}{\sqrt{2}}(Q^{+i}_1-iQ^{+i}_2)~.
\eea
As a result, the action (\ref{sq}) takes the form
\bea
\label{sqq}
S_2
&=&-
\int d\zeta^{(-4)} \Big\{
\bar{\c}^{++}(\Box +2W\bar{W}-\frac{1}{\sqrt{2}}((D^{+\a}W){\cal D}^-_{\a}+
(\bar{D}^{+\ad}W)\bar{\cal D}^-_{\ad})\c^{++}+
\nonumber\\
&+&\bar{\eta}^{+i}(D^{++}-\sqrt{2}V^{++})\eta^+_i
+\hf v^{++}_3\Box v^{++}_3+\hf Q^{+a}_3D^{++}Q^+_{a3}+
\nonumber\\&& + 
i\c^{++}\sqrt{2}q^{+a}\bar{\eta}^+_a+
i\bar{\c}^{++}\sqrt{2}q^{+a}\eta^+_a)
\Big\}
\,. \nonumber \eea
It is convenient to rewrite the action (\ref{sq}) as
\bea
\label{veract}
S_2=S_0+V+\ldots~,
\eea
where
\bea
\label{s0n}
S_0
&=&-
\int d\zeta^{(-4)} \Big\{
\bar{\c}^{++}(\Box +2W\bar{W}-\frac{1}{\sqrt{2}}((D^{+\a}W){\cal D}^-_{\a}+
(\bar{D}^{+\ad}W)\bar{\cal D}^-_{\ad})\c^{++}+
\nonumber\\
&+&\bar{\eta}^{+i}(D^{++}-\sqrt{2}V^{++})\eta^+_i\Big\}
\eea
is the part that is used for defining propagators, and
\bea
\label{ver}
V&=&-i\sqrt{2}(\c^{++}q^{+a}\bar{\eta}^+_a+
\bar{\c}^{++}q^{+a}\eta^+_a)
\eea
is the interaction term.
Dots in (\ref{sqq}) stand for the quadratic action of the decoupling
fields $Q^+_{i3},v^{++}_3$ which do not contribute to quantum corrections. 

It is worth mentioning that for calculation of the effective action in the pure $\N=2$ SYM 
sector one must also take into account the third ghosts \cite{bko}. Contributions 
of the third ghosts to effective action will be considered in Section 4. 
They do not contribute to the hypermultiplet-dependent quantum corrections 
computed in Section 3. 

As the next step we introduce the operator
\bea
\label{covboxw}
\covbox_w=\Box +2W\bar{W}-\frac{1}{\sqrt{2}}((D^{+\a}W){\cal D}^-_{\a}+
(\bar{D}^{+\ad}W)\bar{\cal D}^-_{\ad})
\eea
which is obtained from $\covbox$ (\ref{box0}) by retaining in the latter only 
the superfields $W,\bar{W}$ associated with the unbroken $u(1)$ and putting 
them  on shell. Note that the action (\ref{s0n}) is manifestly analytic like (\ref{S21}). 
As we shall see, while calculating the hypermultiplet-dependent 
contributions it suffices to consider only that part of $\covbox_w$
which does not contain derivatives of the background superfields $W,\bar{W}$. 
In what follows we shall use the notation $\covbox_w=\covbox$, with hoping that 
this will not give rise to any confusion. 

The background-dependent propagators for the superfields $\c^{++},
\bar{\c}^{++},\bar{\eta}^{+\a},\eta^+_b$ are then given by the expressions
\cite{bko}
\bea\label{koz}
<\bar{\c}^{++}(z_1,u_1)\,\c^{++}(z_2,u_2)> &=& - 
\frac{i}{2\stackrel{\frown}{\Box}(u_1)\stackrel{\frown}{\Box}(u_2)}
\nonumber\\&\times& {({\cal D}_1^+)^4 ({\cal D}_2^+)^4} \Big\{
\delta^{12}(z_1-z_2) (D^{--}_2)^2\delta^{(-2,2)}(u_1,u_2) \Big\},
\nonumber\\
<\bar{\eta}^{+a}(z_1,u_1)\eta^+_b(z_2,u_2)>&=&i\delta^a_b
\frac{1}{(u^+_1u^+_2)^3} \frac{({\cal D}^+_1)^4 ({\cal D}^+_2)^4}
{\stackrel{\frown}{\Box}}\delta^{12}(z_1-z_2)\ . \eea
It is worth pointing out that the gauge superfield propagator
$<\bar{\c}^{++}\c^{++}>$ can be written in two equivalent
forms \cite{Nguen,GIKOS3}. The form (\ref{koz}) is advantageous in
that it is analytic with respect to both superspace arguments.
  
The exact one-loop non-holomorphic effective $W, \bar W$ potential in the model under
consideration was found in \cite{dine,bbk,bbkl}. The hypermultiplet-dependent part of 
the effective action was restored in the recent paper \cite{BI}. It was shown there 
by an algebraic analysis that the following expression
\bea
\int d^{12}z\, [{\cal H}(W,\bar{W})+{\cal L}_q(W,\bar{W},q^+)] \label{full1}
\eea
is invariant under the hidden $\N=2$ supersymmetry transformations (\ref{hidden})
(and hence under $\N=4$ supersymmetry), if its hypermultiplet-dependent
part has the form
\bea
\label{condsym}
{\cal L}_q(W,\bar{W},q^+)=\frac{1}{(4\pi)^2}\sum_{n=1}^{\infty}c_n
\left(\frac{q^{+a}q^-_a}{W\bar{W}}\right)^n~,
\eea
with
\bea
\label{coef}
c_n=\frac{(-2)^n}{n^2(n+1)}\,.
\eea
This sum converges to
\bea
{\cal L}_q(X)&=&\frac{1}{(4\pi)^2}
\frac{1}{X}[X({\rm Li}_2(X)-1)-(1-X)\ln(1-X)], \quad X \equiv
-2\frac{q^{+a}q^-_a}{W\bar{W}}~,\label{condsym11}
\eea
where ${\rm Li}_2(X)$ is Euler dilogarithm \cite{Erd}.
Our main purpose will be to show
that the expression (\ref{condsym}), (\ref{condsym11}) can be directly
derived from the quantum harmonic formalism by summing up the
appropriate sequence of the one-loop harmonic
supergraphs.
\setcounter{equation}{0}

\section{Hypermultiplet-dependent contributions to effective action}

The one-loop hypermultiplet-dependent contributions to the
effective action are presented by the following infinite sequence of supergraphs:

\vspace*{2mm}

\hspace*{.5cm}
\Lengthunit=1.1cm
\GRAPH(hsize=3){\mov(0,-.5){
\halfcirc(1.0)[u]\halfwavecirc(1.0)[d]
\mov(-.6,0){\lin(-.5,0)}\mov(.4,0){\lin(.5,0)}}
\ind(18,-5){+}\ind(12,-3){2}\ind(-11,-3){1}
}
\GRAPH(hsize=3){\lin(-.7,.7)\wavelin(1.1,0)\lin(0,-1)
\mov(1,-1){\lin(.7,-.7)\lin(0,1)\wavelin(-1.1,0)}\ind(19,-5){+}
\mov(.9,0){\lin(.7,.7)}\mov(-.2,-1){\lin(-.7,-.7)}
\ind(16,5){3}\ind(16,-15){4}\ind(-12,5){2}\ind(-12,-15){1}
}
\hspace*{.5cm}
\GRAPH(hsize=3){\lin(-.7,.7)\wavelin(.6,0)\lin(0,-1)
\mov(.5,0){\lin(0,1)\lin(.45,0)}
\mov(1,-1){\lin(.7,-.7)\wavelin(0,1)\lin(-.5,0)\ind(13,4){+}
\mov(-.6,0){\lin(0,-1)\wavelin(-.6,0)}}
\mov(.9,0){\lin(.7,.7)}\mov(-.2,-1){\lin(-.7,-.7)}
\ind(16,5){4}\ind(16,-15){5}\ind(-12,5){2}\ind(-12,-15){1}
\ind(4,8){3}\ind(4,-18){6}
}
\hspace*{.5cm}
\GRAPH(hsize=3){\mov(.1,0){\lin(-.7,.7)\wavelin(.6,0)}
\lin(0,-.5)\mov(0,-.5){\wavelin(0,-.5)\lin(-.7,0)}
\mov(.5,0){\lin(0,1)\lin(.5,0)}
\mov(1,-1){\lin(.7,-.7)\lin(0,.5)\wavelin(-.5,0)
\mov(-.6,0){\lin(0,-1)\lin(-.6,0)}}
\mov(.9,-.5){\wavelin(0,.5)\lin(.7,0)}
\mov(.8,0){\lin(.7,.7)}\mov(-.3,-1){\lin(-.7,-.7)}
\ind(16,5){4}\ind(16,-15){6}\ind(-12,5){2}\ind(-12,-15){8}
\ind(4,8){3}\ind(4,-18){7}\ind(17,-5){5}\ind(-15,-5){1}
\ind(21,-5){+}\ind(27,-5){\ldots}
}
\vspace*{2mm}

\noindent Here the external and internal straight lines denote, respectively,
external background hypermultiplets and quantum hypermultiplet propagators,
and the wavy lines stand for the gauge superfield propagators. The numbers
1,2,$\ldots$ mark harmonic arguments of the external hypermultiplets and vertices. We
note that the whole dependence of the corresponding contributions on the external gauge
superfield strengths $W,\bar{W}$ is already accounted for by the
background-dependent propagators (\ref{koz}). This prevents one
from considering `mixed' supergraphs with both $\N=2$ gauge and
hypermultiplet external legs. It is enough to consider only the supergraphs
shown above.

A generic supergraph with $2n$ external lines can be viewed as a ring
consisting of $n$ links of the form 
$<\bar{\eta}^+\eta^+><\c^{++}\bar{\c}^{++}>$ or $n$ links of the
form $<\eta^+\bar{\eta}^+><\bar{\c}^{++}\c^{++}>$. For
external superfields slowly varying in space-time the total contribution of these 
two kinds of the $2n$-point supergraph is given by the following general expression \bea \label{gnin}
\Gamma_{2n}&=&\frac{4}{n}\int d^{12}z \int du_1\ldots du_{2n}
\frac{1}{\covbox(u_1)\covbox^2(u_2)\ldots \covbox(u_{2n-1})\covbox^2(u_{2n})}
\nonumber\\&\times& \frac{ ({\cal
D}^+(u_1))^4({\cal D}^+(u_2))^4(D^{--}_2)^2\delta^{(2,-2)}(u_2,u_3) ({\cal
D}^+(u_3))^4 ({\cal D}^+(u_4))^4
(D^{--}_4)^2\delta^{(2,-2)}(u_4,u_5)}{(u^+_1u^+_2)^3(u^+_3u^+_4)^3}
\nonumber\\&\times& \frac{({\cal D}^+(u_5))^4 \ldots ({\cal
D}^+(u_{2n-1}))^4({\cal D}^+(u_{2n}))^4(D^{--}_{2n})^2
\delta^{(2,-2)}(u_{2n},u_1)} {(u^+_5u^+_6)^3(u^+_7u^+_8)^3\ldots (u^+_{2n-3}
u^+_{2n-2})^3 (u^+_{2n-1} u^+_{2n})^3} \delta^{12}(z-z')|_{z=z'}
 \nonumber\\ &\times&
q^+_a(u_1)q^{+a}(u_2)q^+_b(u_3)\ldots q^+_c(u_{2n-1})q^{+c}(u_{2n})~.
\eea
Hereafter, the symbol $\covbox$ stands for the operator defined by eq. (\ref{covboxw}). 
Since our aim consists in calculation of contribution depending only on $q^{+},W,\bar{W}$ 
but not on their derivatives we can omit all the derivative-depending terms in $\covbox$.
The factor $4/n$ has the following origin.
The contribution from the ring-type supergraph composed of $n$ repeating links
\\ 
$<\bar{\eta}^+\eta^+><\c^{++}\bar{\c}^{++}>$ appears with the symmetry
factor $2/n$. The same factor $2/n$ arises from the supergraph 
composed of $n$ repeating links $<\eta^+\bar{\eta}^+><\bar{\c}^{++}\c^{++}>$. Further, 
each vertex brings the factor $-i$, every  $<\eta^+\bar{\eta}^+>$ and $<\bar{\c}^{++}\c^{++}>$ 
propagators contribute the factors $i$ and $i/2$, respectively 
(hence total of $n$ links contributes $2^{-n}$). Any vertex also carries 
the coefficient $\sqrt{2}$. This leads to the total factor $2^n$. Putting all 
these contributions together, we obtain just the coefficient $4/n$.

We calculate the expression (\ref{gnin}) in the following way. It contains
harmonic integrals of products of the expressions $({\cal D}^+(u_{2k}))^4
(D^{--}_{2k})^2\delta^{(2,-2)}(u_{2k},u_{2k+1})({\cal
D}^+(u_{2k+1}))^4\,$, $k=1,2\ldots n\,$ (with $u_{2n+1}\equiv u_1$).
In every such term we use the identity
$(D^{--}_{2k})^2\delta^{(2,-2)}(u_{2k},u_{2k+1})=
(D^{--}_{2k+1})^2\delta^{(-2,2)}(u_{2k},u_{2k+1})$. Then, integrating by
parts, we throw
$(D^{--}_{2k+1})^2$ on $({\cal D}^+(u_{2k+1}))^4$, integrate over
$u_{2k+1}$ using the harmonic delta function and eventually
obtain the block\\ $({\cal D}^+(u_{2k}))^4 (D^{--}_{2k})^2({\cal
D}^+(u_{2k}))^4$. Due to the identity \cite{bko}
\bea \label{box1}
\stackrel{\frown}{\Box}=-\frac{1}{2}({\cal D}^+)^4(D^{--})^2
\eea
we can replace $({\cal D}^+(u_{2k}))^4(D^{--}_{2k})^2$ by
$-2\covbox_{2k}$ with the operator $\covbox$ given by
(\ref{covboxw}). 
We carry out this replacement for all $k$ except for $k=n$ and, after some
relabelling of harmonics ($u_{2k}$ is relabelled as
$u_{k+1}\,$, $k = 1,2 \ldots n$), we obtain
\bea \label{gnalt}
\Gamma_{2n}&=&\frac{4(-1)^{n-1}2^n}{2n}\int d^{12}z \int du_1 du_2\ldots du_n
du_{n+1}
\frac{1}{\covbox(u_1)\covbox(u_2)\ldots \covbox(u_{n-1})\covbox^2(u_n)}
\nonumber\\&\times&
\frac{({\cal D}^+(u_1))^4 ({\cal D}^+(u_2))^4 \ldots ({\cal
D}^+(u_n))^4({\cal D}^+(u_{n+1}))^4 }
{(u^+_1u^+_2)^3(u^+_2u^+_3)^3\ldots(u^+_n u^+_{n+1})^3}
(D^{--}_{n+1})^2\delta^{(2,-2)}(u_{n+1},u_1)\nonumber\\ &\times&
\delta^{12}(z-z')|_{z=z'} q^+_a(u_1)q^{+a}(u_2)q^+_b(u_2)\ldots
q^+_c(u_n)q^{+c}(u_{n+1})~.
\eea
This general relation is applicable for any $n$.

For $n>1$ one can achieve a further simplification in
(\ref{gnalt}). We make use of the identity (\ref{box1}) one time
more, relabel for convenience the indices ($c$ for $a$, $a$ for $b$, etc.)
and integrate over $u_{n+1}$, after which (\ref{gnalt}) for $n>1$ is reduced
to the expression
\bea
\label{gnm}
\Gamma_{2n}&=&\frac{4(-1)^n2^n}{n}\int d^{12}z
\int du_1 du_2\ldots du_n
\frac{1}{\covbox(u_1)\covbox(u_2)\ldots \covbox(u_{n-1})\covbox(u_n)}
\nonumber\\&\times&
\frac{({\cal D}^+(u_1))^4 ({\cal D}^+(u_2))^4\ldots
({\cal D}^+(u_n))^4}
{(u^+_1u^+_2)^3(u^+_2u^+_3)^3\ldots(u^+_n u^+_1)^3}
\delta^{12}(z-z')|_{z=z'}
\nonumber\\ &\times&
q^{+a}(u_1)q^+_b(u_1)q^{+b}(u_2)\ldots q^+_a(u_n)~.
\eea
Notice that we could arrive at the same result by using the
alternative expression
for the propagator of $\N=2$ gauge superfield:
\bea
<v^{++}_{1A}(z_1,u_1)\,v^{++}_{B1}(z_2,u_2)>
&=&  \delta_{AB}
\frac{i}{\stackrel{\frown}{\Box}_w(u_1)}
{({\cal D}_1^+)^4}
\Big\{ \delta^{12}(z_1-z_2) \delta^{(-2,2)}(u_1,u_2)
\Big\}\,.
\eea
This propagator, unlike (\ref{koz}) which is manifestly analytic in its both arguments,
displays the {\it manifest} analyticity only with respect to $(z_1, u_1)\,$. The
analyticity with respect to the second argument follows only with taking
into account the harmonic delta function \cite{GIKOS3}.

It is appropriate here to note that the expression (\ref{gnm})
obtained with using this `short' form of propagator is ill-defined for $n=1$. So
the $n=1$ case needs a special treatment.
Indeed, on the one hand, at $n=1$ the numerator in (\ref{gnm}) is proportional to
$({\cal D}^+(u_1))^4\delta^{12}(z-z')|_{z=z'}$ which is zero.
On the other hand, the denominator of (\ref{gnm}) for $n=1$ is
$(u^+_1u^+_1)^3 = 0$. Therefore, we encounter a harmonic singularity
of the form $\frac{0}{0}$. To obtain the correct result for the $n=1$ contribution
we should proceed from the expression (\ref{gnalt}) without reducing it to (\ref{gnm}).
This amounts to keeping the $\N=2$ gauge superfield propagator
in the manifestly analytic `long' form (\ref{koz}). The necessity to
use just the 'long' form
of the propagator to avoid coincident harmonic singularities was
emphasized in \cite{Nguen}.

The expression (\ref{gnalt}) for $n=1$ is
\bea
\label{gnalt1}
\Gamma_2&=&4\int d^{12}z
\int du_1 du_2
\frac{1}{\covbox_1\covbox_2^2}
\frac{({\cal D}^+(u_1))^4 ({\cal D}^+(u_2))^4}
{(u^+_1u^+_2)^3}
(D^{--}_2)^2\delta^{(2,-2)}(u_2,u_1)\nonumber\\ &\times&
\delta^{12}(z-z')|_{z=z'}
q^+_a(u_1)q^{+a}(u_2)~.
\eea
Using the relation \cite{bko}
\bea
\label{point}
({\cal D}^+(u))^4 ({\cal D}^+(u'))^4\delta^8(\q-\q')|_{\q=\q'}=
(u^+u^{'+})^4\,,
\eea
one finds
\bea
\label{g2}
\Gamma_2&=&4\int d^{12}z\int du_1 du_2(u^+_1u^+_2)
(D^{--}_2)^2\delta^{(2,-2)}(u_2,u_1)\nonumber\\ &\times&
\frac{1}{\covbox_1\covbox_2^2}\delta^4(x-x')|_{x=x'}
q^+_a(u_1)q^{+a}(u_2)~.
\eea
Then one can take $(D^{--})^2$ off a delta function. Since we are
interested in contributions depending only on $q,W,\bar{W}$ but not
on their derivatives, at this step we can omit all the derivatives of
the background strengths and replace $\covbox$ by $\Box+2W\bar{W}$.
As the result, the
harmonic delta function becomes free of derivatives, and after
integration over one set of harmonics (e.g. over $u_2$) one obtains
\bea
\label{g3}
\Gamma_2&=&-2\int d^{12}z\int du_1
\frac{1}{(\Box+2W\bar{W})^3}\delta^4(x-x')|_{x=x'}
q^{+a}(u_1)q^-_a(u_1)~.
\eea
After performing Fourier
transformation this correction can be written as
\bea
\label{g4}
\Gamma_2&=&-2\int d^{12}z\int du_1 q^{+a}(u_1)q^-_a(u_1)~
\int \frac{d^4k}{(2\pi)^4}\frac{1}{(-k^2+2W\bar{W})^3}\,.
\eea
Doing the momenta integral, we finally cast it in the form
\bea
\label{g5}
\Gamma_2&=&-2\int d^{12}z\int du_1 q^{+a}(u_1)q^-_a(u_1)~
\frac{1}{2(4\pi)^2 W\bar{W}}\,.
\eea
This expression precisely matches with the general result
(\ref{condsym}) -- (\ref{condsym11})
obtained by invoking `hidden' $\N=2$ supersymmetry \cite{BI}.

Let us turn to the generic case of $n>1$.
Since we are interested in
contributions which do not depend on space-time derivatives of
background hypermultiplets, we can regard the latter
to be space-time constants.

To simplify the expression (\ref{gnm}), we proceed as follows.
First, we represent $q^+_b(u_1)$ in (\ref{gnm}) as
$q^+_b(u_1)=D^{++}_1q^-_b(u_1)$ (this is possible since $q^+_b$
sits on its free mass shell)
\bea
\label{gn1m}
\Gamma_{2n}&=&\frac{4(-1)^n}{2^n n}\int d^{12}z
\int du_1\ldots du_n
\frac{1}{\covbox(u_1)\covbox(u_2)\ldots \covbox(u_{n-1})\covbox(u_n)}
\nonumber\\&\times&
\frac{({\cal D}^+(u_1))^4({\cal D}^+(u_2))^4\ldots ({\cal
D}^+(u_n))^4}
{(u^+_1u^+_2)^3(u^+_2u^+_3)^3\ldots(u^+_n u^+_1)^3}\delta^{12}(z-z')|_{z=z'}
\nonumber\\ &\times&
q^{+a}(u_1)D^{++}_1q^-_b(u_1)q^{+b}(u_2)\ldots q^+_a(u_n)\,.
\eea
Then, integrating by parts and using the mass-shell condition $D^{++}_1q^{+a}(u_1)=0$,
we throw the harmonic derivative $D^{++}_1$ on the harmonic factor
\bea
\label{frac}
\frac{1}{(u^+_1u^+_2)^3(u^+_2u^+_3)^3\ldots(u^+_n u^+_1)^3}\,.
\eea
We note that acting of $D^{++}_1$ on $\covbox(u_1)$ would lead only to 
terms which are proportional to derivatives of the background superfield strengths
and so are irrelevant for our purpose of calculating the leading contributions.
The result of acting of $D^{++}_1$ on (\ref{frac}) is
\bea
\label{fa}
& &D^{++}_1\frac{1}{(u^+_1u^+_2)^3(u^+_2u^+_3)^3\ldots(u^+_n u^+_1)^3}=
-\frac{1}{2}
[(D^{--}_2)^2\delta^{(2,-2)}(u_1,u_2)\frac{1}{(u^+_1u^+_n)^3}+
(u_2\leftrightarrow u_n)]
\nonumber\\&\times&
\frac{1}{(u^+_2u^+_3)^3\ldots(u^+_{n-1} u^+_n)^3}\,.
\eea
After substituting (\ref{fa}) in (\ref{gn1m}) we obtain
\bea
\label{gn2m}
\Gamma_{2n}&=&\frac{4(-1)^n2^n}{n}\int d^{12}z
\int du_1\ldots du_n
\frac{1}{\covbox(u_1)\covbox(u_2)\ldots \covbox(u_{n-1})\covbox(u_n)}
\nonumber\\&\times&
\frac{1}{2}\,
[(D^{--}_2)^2\delta^{(2,-2)}(u_1,u_2)\frac{1}{(u^+_1u^+_n)^3}+(u_2\to
u_n)]\frac{1}{(u^+_2u^+_3)^3\ldots(u^+_{n-1} u^+_n)^3}
\nonumber\\&\times&
({\cal D}^+(u_1))^4({\cal D}^+(u_2))^4\ldots ({\cal D}^+(u_n))^4
\delta^{12}(z-z')|_{z=z'}
\nonumber\\ &\times&
q^{+a}(u_1)q^-_b(u_1)q^{+b}(u_2)\ldots q^{+d}(u_n)q^+_a(u_n)~.
\eea
Then we take the factor $(D^{--}_2)^2$ off the delta function.
It is easy to see that this factor can give a non-vanishing result only when hitting $({\cal
D}^+(u_2))^4$.
Then we relabel $u_2\leftrightarrow u_n$ in the second
term in the square bracket in (\ref{gn2m}), after which it becomes identical to
the first one. Doing integral over $u_1$, we arrive at the following
intermediate expression for $\Gamma_{2n}$
\bea
\label{gn3m}
\Gamma_{2n}&=&-\frac{4(-1)^n2^n}{n}\int d^{12}z
\int du_2\ldots du_n
\frac{1}{\covbox(u_2)\covbox(u_2)\ldots \covbox(u_{n-1})\covbox(u_n)}
\nonumber\\&\times&
\frac{1}{(u^+_2u^+_3)^3\ldots(u^+_{n-1} u^+_n)^3(u^+_n u^+_2)^3}
\nonumber\\&\times&
({\cal D}^+(u_2))^4[(D^{--}_2)^2({\cal D}^+(u_2))^4]
\ldots ({\cal D}^+(u_n))^4\delta^{12}(z-z')|_{z=z'}
\nonumber\\ &\times&
q^{+a}(u_2)q^-_b(u_2)q^{+b}(u_2)q^+_c(u_2)\ldots q^{+d}(u_n)q^+_a(u_n)~.
\eea

At the next step we repeat above procedure for $q^+_c(u_2)$ in (\ref{gn3m}),
i.e. represent it in the form $q^+_c(u_2)=D^{++}_2q^-_c(u_2)$  and integrate
by parts with respect to $D^{++}_2$.
Performing the same manipulations as in deriving (\ref{gn3m}), we bring
$\Gamma_{2n}$ into the form
\bea \label{gn3a}
\Gamma_{2n}&=&(-1)^2\frac{4(-1)^n2^n}{n}\int d^{12}z \int du_3\ldots du_n
\frac{1}{\covbox(u_2)\covbox(u_2)\ldots \covbox(u_{n-1})\covbox(u_n)}
\nonumber\\&\times&
\frac{1}{(u^+_3u^+_4)^3\ldots(u^+_{n-1} u^+_n)^3(u^+_n u^+_3)^3}
\nonumber\\&\times& ({\cal D}^+(u_3))^4[(D^{--}_3)^2({\cal
D}^+(u_3))^4]^2({\cal D}^+(u_4))^4 \ldots ({\cal
D}^+(u_n))^4\delta^{12}(z-z')|_{z=z'} \nonumber\\ &\times&
q^{+a}(u_3)q^-_b(u_3)q^{+b}(u_3)q^-_c(u_3)\ldots q^{+d}(u_n)q^+_a(u_n)\,.
\eea
Note that $D^{++}_2$, when hitting $({\cal D}^+(u_2))^4[(D^{--}_2)^2({\cal
D}^+(u_2))^4]$, gives rise to the structures like
$({\cD}^+(u))^4({\cD}^+(u))^3=0$. After $k$ analogous steps, we have
\bea \label{gn3b}
\Gamma_{2n}&=&(-1)^k\frac{4(-1)^n2^n}{n}\int d^{12}z \int du_{n-k}\ldots du_n
\frac{1}{\covbox^k(u_{k+1})
\ldots \covbox(u_{n-1})\covbox(u_n)}\nonumber\\&\times&
\frac{1}{(u^+_{k+1}u^+_{k+2})^3\ldots(u^+_{n-1} u^+_n)^3(u^+_n u^+_{k+1})^3}
\nonumber\\&\times& ({\cal D}^+(u_k))^4[(D^{--}_k)^2({\cal
D}^+(u_k))^4]^k({\cal D}^+(u_{k+1}))^4 \ldots ({\cal
D}^+(u_n))^4\delta^{12}(z-z')|_{z=z'}  \nonumber\\ &\times&
q^{+a}(u_{n-k})q^-_{a_1}(u_{n-k})q^{+a_1}(u_{n-k})q^-_{a_2}(u_{n-k})\ldots
q^{+a_k}(u_{n-k})\ldots q^{+d}(u_n)q^+_a(u_n)\,.
\eea

Repeating the same procedure further, we can reduce the harmonic integral to that
over three sets of harmonics, $u,\, u'$ and $u''$:
\bea \label{gn4}
\Gamma_{2n}&=&\frac{4(-1)^{2n-3}2^n}{n}\int d^{12}z \int du du' du''
\frac{1}{\covbox_u^{n-2}\covbox_{u'}\covbox_{u''}}\nonumber\\&\times&
[({\cal D}^+(u))^4(D^{--})^2]^{n-3}({\cal D}^+(u))^4
({\cal D}^+(u'))^4({\cal D}^+(u''))^4\delta^{12}(z-z')|_{z=z'}
\nonumber\\ &\times&
\frac{1}{(u^+u^{'+})^3(u^{'+}u^{''+})^3(u^{''+}u^+)^3}
q^{+a}(u)q^-_b(u)q^{+b}(u)\ldots \nonumber\\&\times&
q^{+e}(u)q^+_c(u)
q^{+c}(u')q^+_d(u')q^{+d}(u'')q^+_a(u'')~.
\eea
Then, using (\ref{box1}), we can reduce the number of
spinor derivatives in the numerator by the relation
\bea
\label{proj}
[({\cal D}^+(u))^4(D^{--})^2]^{n-3}=(-2)^{n-3}
\covbox_u^{n-3}~.
\eea
After this we once again effect the previous procedure by writing
$q^+_c(u)=D^{++}_uq^-_c(u)$ and throwing $D^{++}_u$
on the harmonic factor. Repeating the same steps as above and using at the last step the
relation (\ref{proj}), we can perform the $u'$-integration, thus arriving at the
expression
\bea
\label{ct61a1}
\Gamma_{2n}&=&-(-2)^n2^n\int d^{12}z
\int du du''
\frac{1}{\covbox_u^n\covbox_{u'}\covbox_{u''}}
\frac{({\cal D}^+(u))^4 ({\cal D}^+(u''))^4}
{(u^+u^{''+})^6}\delta^{12}(z-z')|_{z=z'}
\nonumber\\ &\times&
q^{+a}(u)q^-_b(u)q^{+b}(u)\ldots 
q^{+e}(u)q^-_c(u)
q^{+c}(u)q^+_d(u)q^{+d}(u'')q^+_a(u'')~.
\eea
Now we suppress all the derivative-depending terms in $\covbox$
by replacing $\covbox$ with $\Box+2W\bar{W}$. After this we use 
the identity (\ref{point}). We find
\bea
\label{ct61a2}
\Gamma_{2n}&=&-\frac{(-2)^n2^n}{n}\int d^{12}z
\int du du''
\frac{1}{(\Box+2W\bar{W})^{n+2}}\delta^4(x-x')|_{x=x'}\nonumber\\&\times&
\frac{1}{(u^+u^{''+})^2}
q^{+a}(u)q^-_b(u)q^{+b}(u)\ldots q^+_c(u) q^{+c}(u'')q^+_a(u'')\,.
\eea
The last step is to rewrite $q^+_a(u'')=D^{++}_{u''}q^-_a(u'')$ and
to throw $D^{++}_{u''}$  on $1/(u^+u^{''+})^2\,$, which gives
$$
D^{++}_{u''}\frac{1}{(u^+u^{''+})^2}=D^{--}_{u''}
\delta^{(2,-2)}(u'',u)=-D^{--}_u\delta^{(0,0)}(u'',u)\,.
$$
Then we throw $D^{--}_u$ on $q^+_c(u)$, integrate over $u''$ and
perform the Fourier transformation. The expression
for the contribution of $2n$-th order in hypermultiplets which we
obtain at this step is as follows
\bea
\label{gn5}
\Gamma_{2n}&=&\frac{(-2)^n2^n}{n}\int d^{12}z
\int du \int\frac{d^4k}{(2\pi)^4}
\frac{1}{(-k^2+2W\bar{W})^{n+2}}
\nonumber\\ &\times&
q^{+a}(u)q^-_b(u)q^{+b}(u)\ldots q^{+c}(u)q^-_a(u)~.
\eea
Since (in Minkowski space)
\bea
\int\frac{d^4k}{(2\pi)^4}\frac{1}{(-k^2+2W\bar{W})^{n+2}}=
\frac{1}{(4\pi)^2}
\frac{\Gamma(n)}{\Gamma(n+2)}\frac{1}{(2W\bar{W})^n}\,,
\eea
we find the final expression in the form
\bea
\label{gn6}
\Gamma_{2n}&=&\frac{(-2)^n}{n^2(n+1)}
\frac{1}{(4\pi)^2}\int d^{12}z
\frac{1}{(W\bar{W})^n}
(q^{+a}(u)q^-_a(u))^n\,.
\eea
This expression precisely coincides with the result
(\ref{condsym}) - (\ref{condsym11}) obtained from the requirement of
${\cal N}=4$ supersymmetry.
This expression can be represented as
\bea
\label{gn7}
\Gamma_{2n}&=&\frac{1}{n^2(n+1)}
\frac{1}{(4\pi)^2}\int  d^{12} z X^n\,,
\eea
where
\bea
X=-2\frac{q^{+a}q^-_a}{W\bar{W}}
\eea
(note that in the central basis of $\N=2$ harmonic superspace the hypermultiplet superfields 
are expressed on shell as $q^{\pm\,a} =q^{ia}(z)u^{\pm}_i$, and so $X$ and $\Gamma_{2n}$ do not depend on
harmonics \cite{BI}). The result for $n=1$, eq. (\ref{g5}),
is also incorporated by the general expression (\ref{gn7}).

Thus the total one-loop effective action in the hypermultiplet sector
explicitly computed using the harmonic supergraph techniques
is given by the expression (\ref{condsym}) - (\ref{condsym11})
anticipated on the $\N = 4$ supersymmetry ground in \cite{BI}.
Combined with the non-holomorphic effective potential,
it  provides the exact $\N=4$ supersymmetric low-energy
$\N =4$ SYM effective action (\ref{full1}) \cite{BI}.
\setcounter{equation}{0}

\section{A manifestly $\N=2$ supersymmetric calculation of the
non-holomorphic effective potential of $W, \bar W$}
Here we demonstrate that the techniques developed in section 3
are equally applicable to calculating the non-holomorphic
effective potential (\ref{nonhol}), with taking into account 
some specific features of this problem. The calculation
is carried out entirely within the $\N =2$ harmonic superfield
formalism, with preserving manifest
$\N =2$ supersymmetry.\footnote{A different method of
manifestly ${\cal N}=2$ supersymmetric calculation of the potential
(\ref{nonhol}) was developed in \cite{kuz} using proper-time techniques.}

To compute the one-loop low-energy effective action of $W, \bar W$ in the
$\N =2$ harmonic background field method, we again make the
splitting $V^{++}\to V^{++}+gv^{++}$ in (\ref{2}), with $V^{++}$ being a
background field and $v^{++}$ a quantum one. This part of the full 
effective action of the theory under consideration is determined by the
contributions from the pure $\N=2$ SYM sector, hypermultiplet sector,
Faddeev-Popov (FP) ghosts and third ghosts.
However, the contributions from the
hypermultiplet and FP ghost sectors in the one-loop approximation are known to
exactly cancel each other \cite{bbka} (see also
\cite{bbkl}). The reason for this cancellation is as follows. It was shown in
\cite{bko} that the action of FP ghosts (with choosing
the background-gauge covariant version of Fermi gauge
for the quantum superfields $v^{++}$) has the same form as
the action of massless hypermultiplet in the adjoint representation. The
$\N=4$ SYM theory is $\N =2$ SYM plus a hypermultiplet in adjoint
representation. So the contribution of FP ghosts to $\N=4$ SYM one-loop
effective  action is equal to that of hypermultiplets taken with the
opposite sign (hypermultiplets are bosons, while FP ghosts are fermions).
Therefore, that part of the full one-loop effective action which depends only on $\N=2$ gauge
superfields is constituted by contributions from the pure $\N=2$ SYM sector and 
the third ghosts sector.

The general form of this part of the one-loop effective action was found in \cite{bbk,bbkl,bbka}:
\bea
\label{g1}
\Gamma^{(1)}=\frac{i}{2} Tr_{(2,2)} \ln\covbox - \frac{i}{2}
Tr_{(4,0)}\ln\covbox~.
\eea
Here the symbol $Tr$ means both the functional trace and the one with respect to
matrix indices. The operator $\covbox$ is given by the on-shell expression (\ref{covboxw}),  
with $W,\bar{W}$ being matrices which take values in the Cartan subalgebra of $su(2)$. 
The exact matrix structure of this operator  will be explored further.
For our calculation in this Section it will be important that $\covbox$ includes both the $W\bar{W}$ 
and derivative-depending terms. First contribution in (\ref{g1}) originates from the $\N=2$ SYM sector, and
the second one is determined by the third ghosts $\rho^{(+4)},\sigma$ 
\cite{bbkl}. The propagator of these ghosts reads \cite{bbkl}
\bea
\label{ptg}
G^{(4,0)}(1,2)=<\rho^{(+4)}(1)\sigma(2)> = (D^{++})^2_1 G^{(0,0)}(1,2)~,
\eea
where $G^{(0,0)}$ is the propagator of an uncharged analytic superfield $\omega$ \cite{bbkl}.
The precise form of $G^{(0,0)}$ will be of no need for our further purposes.

Let us consider a theory of uncharged analytic superfield  $\sigma$ with the action 
\bea
\label{alts}
-\hf {\rm tr}\int d\zeta^{(-4)} \sigma (D^{++})^2 \covbox\sigma~.
\eea
This theory leads to the same Green function (\ref{ptg}) and hence to
the same contribution to the effective action (\ref{g1}). 
Therefore the action (\ref{alts}) can be treated as an alternative
form of the action of third ghosts. It will prove to be more 
convenient for our purposes. Hereafter, the symbol `$\mbox{tr}$' denotes trace over 
matrix indices only.

Thus the quadratic action of quantum superfields is given by
\bea
-\hf {\rm tr}\int d\zeta^{(-4)} v^{++}\covbox v^{++}
-\hf {\rm tr}\int d\zeta^{(-4)} \sigma (D^{++})^2\covbox \sigma~.
\label{2aq}
\eea
As the next important step, we separate $v^{++}$ in the two orthogonal projections
\bea
& &v^{++}=v^{++}_T + D^{++}\xi~,\label{tra1} \\ 
& &D^{++}v^{++}_T=0~.\label{Transv}
\eea
The corresponding projecting operator is given in \cite{kuz} and it is a 
covariantization of the analogous operator introduced in \cite{GIKOS,GIKOS3}. The covariantized 
projecting operator $\P^{(2,2)}_T(1,2)$ is defined by the expression
\bea
\label{pro}
v^{++}_{T}(1) &=& \int d\zeta_2^{(-4)}
\P^{(2,2)}_T(1,2)v^{++}(2)~,
\eea
and preserves the analyticity and the harmonic constraint in (\ref{Transv}).\footnote{I.L.B. is very thankful to
S.M. Kuzenko for clarifying the technical details of calculations in Ref.\cite{kuz}, especially those 
related to the structure and properties of the projection operator $\P^{(2,2)}_T$.} Its explicit form can be 
found in \cite{kuz}.

After splitting $v^{++}$ as in (\ref{tra1}), the action (\ref{2aq}) can be rewritten as
\bea
-\hf {\rm tr}\int d\zeta^{(-4)} v^{++}_{T}\covbox v^{++}_{T}
+\hf {\rm tr}\int d\zeta^{(-4)} \xi (D^{++})^{2}\covbox \xi
-\hf {\rm tr}\int d\zeta^{(-4)} \sigma (D^{++})^{2}\covbox\sigma~.
\eea
We see that $\xi$ is a boson, while $\sigma$ is a fermion and they have the
same actions. Clearly, their contributions to the one-loop effective action are
equal up to the sign and hence cancel each other.
Therefore it is sufficient to consider only the following part of the
quadratic action:
\bea
\label{suff00}
S_2=-\hf{\rm tr} \int d\zeta^{(-4)} v^{++}_{T}
\covbox v^{++}_{T}.
\eea
Further,  quantum fields $v^{++}_T$ with values in $su(2)$
can be written in the form (\ref{exgen10}) where  $\t_i$
are generators of $su(2)$ related to Pauli matrices $\s_i$ by 
(\ref{gen10}). They satisfy the relations (\ref{rela}). 
The background $\N=2$ superfields are expanded as
${\cal W}={\cal W}_i\t_i$.
We calculate the effective action with the background superfields taking values
in the unbroken $u(1)$. This corresponds to the restriction
${\cal W}={\cal W}_3\t_3$. As before, we denote ${\cal W}_3$ by $W$.

The quadratic action of quantum superfields for the given background can be read off 
from (\ref{s0n}), with the background hypermultiplets being put equal to zero:
\bea
\label{suff}
S_2=-\int d\zeta^{(-4)} \Big\{
 \c^{++}_T(\Box +2W\bar{W})\bar{\c}^{++}_T-
\frac{1}{\sqrt{2}}\c^{++}_T(({\cal D}^{+\a}W){\cal D}^-_{\a}+
(\bar{\cal D}^{+\ad}W)\bar{\cal D}^-_{\ad})\bar{\c}^{++}_T
\Big\}
\,.  
\eea
The one-loop effective action corresponding to (\ref{suff}) is \cite{kuz}
\bea
\label{15}
\Gamma^{(1)}=iTr_{(2,2)}^{T}\ln\covbox~,
\eea
where the symbol $Tr^T$ means that the trace is taken over subspace of
the analytic superfields $v^{++}_T$ satisfying the constraint ${\cal
D}^{++}v^{++}_T=0$. Hereafter the operator $\covbox$ acting on 
$\c^{++}_T$ is given by the expression (\ref{covboxw}).
Eq. (\ref{15}) leads to the following expression for 
the one-loop effective action \cite{kuz}:
\bea
\Gamma^{(1)}=i\int
d\z^{(-4)}_1
(\ln \covbox)\P^{(2,2)}_T (1,2)|_{\z_1=\z_2, u_1=u_2}~.
\eea
The projector $\P^{(2,2)}_T$ (\ref{pro}) at the coinciding harmonics
is given by the expression \cite{kuz}
\bea
\P^{(2,2)}_T(1,2)|_{u_1=u_2} &=&  ({\cal D}^+_1)^{4}\delta^{12}(z_1-z_2)~.
\eea
Therefore the one-loop effective action is
\bea
\label{ga1}
\Gamma^{(1)}=i\int
d\z^{(-4)}_1 ({\cal D}^{+}_{1})^{4}
 (\ln \covbox)
\delta^{12}(z_2-z_1)|_{z_1=z_2}~.
\eea

Let us consider the general structure of the quantum corrections.
We substitute the expression $\covbox=\Box+2W\bar{W}+R$ in (\ref{ga1}) 
and obtain
\bea
\label{gamma1}
\Gamma^{(1)}=i\int
d\z^{(-4)}_1 ({\cal D}^{+}_{1})^{4}
 \ln (\Box+2W\bar{W}+R)
\delta^{12}(z_2-z_1)|_{z_2=z_1}~,
\eea
where
\bea
R \equiv -\frac{1}{\sqrt{2}}(({\cal D}^{+\a}W){\cal D}^-_{\a}+
(\bar{\cal D}^{+\ad}{\bar W})\bar{\cal D}^-_{\ad})~.
\eea
Note that the expression (\ref{gamma1}) is free of the harmonic
singularities since it does not contain harmonic delta functions which
could lead, in principle, to appearance of such singularities in harmonic
supergraphs. To find the effective action, we expand (\ref{gamma1})
in power series with respect to R, i.e. spinor derivatives of $W,\bar{W}$
\bea
 \ln (\Box+2W\bar{W}+R)=\ln(\Box+2W\bar{W})+
\sum_{n=1}^{\infty}\frac{(-1)^{n-1}}{n}
\left(\frac{R}{\Box+2W\bar{W}}\right)^n .
\eea
We obtain
\bea
\label{gamma11}
\Gamma^{(1)} &=& i{\rm tr}\int
d\z^{(-4)}_1 ({\cal D}^{+}_{1})^{4}\Big\{\ln(\Box+2W\bar{W})
\nonumber\\& & +
\sum_{n=1}^{\infty}\frac{(-1)^{n-1}}{n}
\left(\frac{1}{\Box+2W\bar{W}}\right)^n R^n\Big\}
\delta^{12}(z_2-z_1)|_{z_2=z_1}~.
\eea
We observe that the first term in (\ref{gamma11}), i.e. the one containing 
$\ln(\Box+2W\bar{W})$ carries only four spinor derivatives coming from  
$({\cal D}^+)^4$. On the other hand, shrinking any $\theta$ loop into a point in 
$\q$-space by the rule (\ref{point}) requires the presence of at least eight 
spinor derivatives. Hence, the first term in (\ref{gamma11})
cannot give contribution to the non-holomorphic effective potential. 

The expansion of $\Gamma^{(1)}$ given by (\ref{gamma11}) with the first term 
omitted can be represented by a sequence of harmonic supergraphs. Each supergraph is 
a ring with $n$ vertices and $n$ internal lines. The internal lines are represented 
by the propagators $(\Box+2W\bar{W})^{-1}\delta^{12}(z_1-z_2)$ defined in
full $\N=2$ superspace. There is one common factor $({\cal D}^+)^4$, and 
at each vertex an external line ${\cal D}^+W$ or $\bar{\cal D}^+\bar{W}$ is attached, with the 
factors ${\cal D}^-$ or $\bar{\cal D}^-$, respectively.
The analyticity of the contribution of any supergraph is guaranteed by the factor $({\cal D}^+)^4$.
Thus the contribution of a generic supergraph contains $({\cal D}^{+}_1)^4$ from (\ref{gamma11}) 
and $n$ ${\cal D}^-$ factors, each being associated with each of $n$ vertices. 

Our aim is to calculate the
contribution to the nonholomorphic effective potential which has the
form $\int d^{12}z du F (W,\bar{W})$, with $F(W,\bar{W})$ being some
function of the background strengths $W,\bar{W}$, but not of their derivatives.
We can rewrite this contribution as an integral over
the analytic harmonic subspace by the rule
\bea
\int d^{12}z du\, F (W,\bar{W})=
\int d\z^{(-4)} ({\cal D}^+)^4 F(W,\bar{W}).
\eea 
This representation shows that the contribution to the nonholomorphic effective 
potential rewritten as an integral over the analytic subspace contains exactly 
two chiral derivatives and two antichiral ones acting on the background $W,\bar{W}$. 
To extract such corrections we must represent all vertices but one as integrals over
$d^{12}z$, and keep one vertex in the form of integral over
$d\z^{(-4)}$. From (\ref{gamma11}) we can derive that the contribution
of an arbitrary supergraph with $n=a+b$ vertices, $a$ vertices
containing ${\cal D}W$ and $b$ vertices containing $\bar{\cal D}\bar{W}$, 
is proportional to the expression
\bea
\Gamma_n\propto\int d\z^{(-4)}({\cal D}W)^a(\bar{\cal D}\bar{W})^b
[{\cal D}^a\bar{\cal D}^b({\cal D}^+)^4\delta^{12}(z-z')|_{z=z'}]~.
\eea
Since for obtaining the contribution to nonholomorphic
effective potential we must keep only the terms containing four supercovariant 
derivatives on $W, \bar W$ (two chiral and two antichiral ones), we are led to 
specialize to the case of $a=b=2$. For $a,b>2$ we would obtain 
contributions which remain to be derivative-dependent after representing 
them as integrals over $d^{12}z$. For $a,b<2$ we would have 
not enough ${\cal D}$-factors to shrink a $\theta$-loop into a point.
Therefore the only terms in the expansion of $\ln\covbox$ which can be relevant 
for our purpose are those of the second order in both ${\cal D}W$ and 
$\bar{\cal D}\bar{W}$. There is only one such term in the expansion
of $\ln\covbox$. It is the $n=4$ term the explicit form of which is
\bea
&& -\left(\frac{1}{\Box+2W\bar{W}}\right)^4\frac{1}{16}\Big[
({\cal D}^{+\a}W){\cal D}^-_{\a}+
(\bar{\cal D}^{+\ad}{\bar W})\bar{\cal D}^-_{\ad}\Big]^4
\nonumber\\& & =
-\frac{6}{16}\left(\frac{1}{\Box+2W\bar{W}}\right)^4
[({\cal D}^{+\a}W){\cal D}^-_{\a}]^2
[(\bar{\cal D}^+_{\ad}{\bar W})\bar{\cal D}^{-\ad}]^2+\ldots~. \label{4order}
\eea
The coefficient $-1/16$ arose due to $-1/4$ from the expansion 
of logarithm and the factor $(1/\sqrt{2})^4$. The coefficient 6 appeared 
from  the binomial expansion 
$\Big[
i({\cal D}^{+\a}W){\cal D}^-_{\a}+
i(\bar{\cal D}^{+\ad}{\bar W})\bar{\cal D}^-_{\ad}\Big]^4$ \\ $= 6
(({\cal D}^{+\a}W){\cal D}^-_{\a})^2
((\bar{\cal D}^{+\ad}{\bar W})\bar{\cal D}^-_{\ad})^2+\ldots$.
Here and in (\ref{4order}) dots stand for terms giving zero trace.

After keeping only the $n=4$ term (\ref{4order}) in $\Gamma^{(1)}$, eq. (\ref{gamma11}), the latter 
can be represented by a four-point supergraph of the form

\vspace*{1mm}

\hspace*{4.5cm}
\Lengthunit=2cm
\Linewidth{.5pt}
\GRAPH(hsize=3){\mov(.1,0){\wavelin(-.5,.5)}
\Linewidth{1.5pt}\wavelin(1.1,0)\wavelin(0,-1)\Linewidth{.5pt}
\mov(1,-1){\wavelin(.5,-.5)\Linewidth{1.5pt}\wavelin(0,1)\wavelin(-1.1,0)}
\Linewidth{.5pt}
\mov(.88,0){\wavelin(.5,.5)}\mov(-.25,-1){\wavelin(-.5,-.5)}
\ind(0,.5){|}\ind(8,-1){-}\ind(7,-10){|}\ind(-2,-9){-}\ind(-3,-1){-}
\ind(0,2.5){{\cal D}^-_{\a}(u)}\ind(13,-1){\bar{\cal D}^{-\ad}(u)}
\ind(5,-13){{\cal D}^-_{\b}(u)}\ind(-9,-9){\bar{\cal D}^{-\bd}(u)}
\ind(-10,-1){({\cal D}^+(u))^4}
\ind(-15,6){{\cal D}^{+\a}(u)W}
\ind(15,6){\bar{\cal D}^+_{\ad}(u){\bar W}}
\ind(15,-16){{\cal D}^{+\b}(u)W}
\ind(-18,-16){\bar{\cal D}^+_{\bd}(u){\bar W}}
}

\vspace*{1mm}

\noindent Here external legs represent derivatives of superfield strengths,
while bold internal lines  stand for the ``free'' propagators 
$\frac{1}{(\Box+2W\bar{W})}\delta^{12}(z_1-z_2)$. The full contribution of this graph 
is given by the following expression:
\bea
\Gamma_4&=&-\frac{6}{16}i\int d\z^{(-4)}_1 d^{12}z_2 d^{12}z_3 d^{12}z_4
({\cal D}^{+\a}(u_1)W(x_1,\q_1)){\cal D}^-_{\a}(u_1)
(\bar{\cal D}^+_{\ad}(u_1){\bar W}(x_2,\q_2))\bar{\cal D}^{-\ad}(u_1)
\nonumber\\&\times&
({\cal D}^{+\b}(u_1)W(x_3,\q_3)){\cal D}^-_{\b}(u_1)
(\bar{\cal D}^+_{\bd}(u_1){\bar W}(x_4,\q_4))
\bar{\cal D}^{-\bd}(u_1)({\cal D}^+(u_1))^4
\nonumber\\&\times&
\frac{1}{(\Box+2W\bar{W})}\delta^{12}(z_1-z_2)
\frac{1}{(\Box+2W\bar{W})}\delta^{12}(z_2-z_3)
\frac{1}{(\Box+2W\bar{W})}\delta^{12}(z_3-z_4)\nonumber\\&\times&
\frac{1}{(\Box+2W\bar{W})}\delta^{12}(z_4-z_1)~.
\eea
We observe that this expression contains only one integral over harmonics
$u_1$ (recall that the harmonic integral is included into $d\z^{(-4)}_1$) 
and does not involve harmonic delta functions due to
the structure of (\ref{gamma1}), (\ref{gamma11}). Therefore it is automatically 
free of harmonic singularities.
Then we anticommute
$\bar{\cal D}^{-\ad}(u)$ with ${\cal D}^-_{\b}(u)$, which produces
factor $-1$. As the result we obtain
\bea
\Gamma_4&=&\frac{6}{16}i\int d\z^{(-4)}_1 d^{12}z_2 d^{12}z_3 d^{12}z_4
({\cal D}^{+\a}(u_1)W(x_1,\q_1))
(\bar{\cal D}^+_{\ad}(u_1){\bar W}(x_2,\q_2))
\nonumber\\&\times&
({\cal D}^{+\b}(u_1)W(x_3,\q_3))
(\bar{\cal D}^+_{\bd}(u_1){\bar W}(x_4,\q_4)){\cal D}^-_{\a}(u_1)
{\cal D}^-_{\b}(u_1)
\bar{\cal D}^{-\ad}(u_1)
\bar{\cal D}^{-\bd}(u_1)({\cal D}^+(u_1))^4
\nonumber\\&\times&
\frac{1}{(\Box+2W\bar{W})}\delta^{12}(z_1-z_2)
\frac{1}{(\Box+2W\bar{W})}\delta^{12}(z_2-z_3)
\frac{1}{(\Box+2W\bar{W})}\delta^{12}(z_3-z_4)\nonumber\\&\times&
\frac{1}{(\Box+2W\bar{W})}\delta^{12}(z_4-z_1)~.
\eea
We then integrate over $\q_3,\q_4$ with the help of delta functions and shrink 
a loop into a point by the rule
\bea
&& \delta^8(\q_1-\q_2)(\bar{\cal D}^{-\ad}(u_1))(\bar{\cal D}^{-\bd}(u_1))
({\cal D}^-_{\a}(u_1))({\cal
D}^-_{\b}(u_1))({\cal D}^+(u_1))^4\delta^8(\q_1-\q_2)
\nonumber \\
&& =
4\epsilon_{\a\b}\epsilon^{\ad\bd}
\delta^8(\q_1-\q_2)~.
\eea
Further we integrate over $\q_2$ and arrive at the expression
\bea
\Gamma_4&=&-\frac{6}{4}i\int d\z^{(-4)}_1  d^4x_2 d^4x_3 d^4x_4
({\cal D}^{+\a}(u_1)W(x_1,\q))({\cal D}^+_{\a}(u_1)W(x_3,\q))
\nonumber\\&\times&
(\bar{\cal D}^+_{\ad}(u_1){\bar W}(x_2,\q))(\bar{\cal
D}^{+\ad}(u_1){\bar W}(x_4,\q))
\nonumber\\&\times&
\frac{1}{(\Box+2W\bar{W})}\delta^4(x_1-x_2)
\frac{1}{(\Box+2W\bar{W})}\delta^4(x_2-x_3)
\frac{1}{(\Box+2W\bar{W})}\delta^4(x_3-x_4)\nonumber\\&\times&
\frac{1}{(\Box+2W\bar{W})}\delta^4(x_4-x_1)~.
\eea
The (non-holomorphic) effective potential by definition is an effective
Lagrangian with the background strengths slowly varying in space-time. 
Therefore we can put
\bea
& &({\cal D}^{+\a}(u_1)W(x_1,\q))({\cal D}^+_{\a}(u_1)W(x_2,\q))
(\bar{\cal D}^+_{\ad}(u_1){\bar W}(x_3,\q))(\bar{\cal
D}^{+\ad}(u_1){\bar W}(x_4,\q))\nonumber\\& &\simeq
({\cal D}^{+\a}(u_1)W(x_1,\q))({\cal D}^+_{\a}(u_1)W(x_1,\q))
(\bar{\cal D}^+_{\ad}(u_1){\bar W}(x_1,\q))(\bar{\cal
D}^{+\ad}(u_1){\bar W}(x_1,\q))~.
\eea
After this we perform Fourier transformation which leads to
\bea
\Gamma_4&=&-\frac{6}{4}i\int d\z^{(-4)}_1
({\cal D}^{+\a}(u_1)W)^2(x,\q)(\bar{\cal D}^+_{\ad}(u_1){\bar W})^2(x,\q)
\int\frac{d^4k}{(2\pi)^4}\frac{1}{(k^2-2W\bar{W})^4}~.
\eea
After doing the loop integration by the rule 
\bea
\int\frac{d^4k}{(2\pi)^4}\frac{1}{(k^2-2W\bar{W})^4}=
\frac{1}{6}\,i\,\frac{1}{(4\pi)^2}\frac{1}{(2W\bar{W})^2}
\eea
we have
\bea
\Gamma_4&=&\frac{1}{16}\int d\z^{(-4)}_1 
({\cal D}^{+\a}(u_1)W)^2(\bar{\cal D}^+_{\ad}(u_1){\bar W})^2
\frac{1}{(4\pi)^2}\frac{1}{(W\bar{W})^2}~.
\eea
Then we note that
\bea
({\cal D}^{+\a}(u_1)W)^2(\bar{\cal D}^+_{\ad}(u_1){\bar W})^2
\frac{1}{(W\bar{W})^2}=16({\cal D}^+(u_1))^4
\left(\ln\frac{W}{\L}\ln\frac{\bar{W}}{\L}\right),
\eea
whence
\bea
\Gamma_4&=&\frac{1}{(4\pi)^2}\int d\z^{(-4)}_1 ({\cal D}^+(u_1))^4
\left(\ln\frac{W}{\L}\ln\frac{\bar{W}}{\L}\right). \label{prefin}
\eea

Finally, we pass to the integration over the full harmonic superspace, 
$\int d\z^{(-4)}_1 ({\cal D}^+(u_1))^4=
\int d^{12}z du_1$, and take into account that 
the $u_1$ integral in (\ref{prefin}) can be taken away because the integrand 
does not depend on harmonics. Hence we finally obtain the complete one-loop 
non-holomorphic effective potential just in the form (1.1) given 
in \cite{dine,bbk,bbkl}.

We conclude that the non-holomorphic potential of $W, \bar W$, like the
hypermultiplet-dependent part of the full effective action, can be
self-consistently derived in the framework of the quantum harmonic 
superspace approach.
\setcounter{equation}{0}

\section{Summary}
To summarize, in this paper we addressed the problem of calculating $\N=4$
supersymmetric low-energy effective action of $\N=4$ $SU(2)$ SYM theory in
the Coulomb branch formulated in harmonic superspace as $\N=2$ SYM theory coupled to
the hypermultiplet in adjoint representation of gauge group. We have
developed a universal procedure of computing both the hypermultiplet-dependent
and purely $\N =2$ SYM parts of the effective action, based
on the covariant harmonic supergraph techniques ensuring a manifest $\N=2$
supersymmetry at every stage of calculation.
The directly computed hypermultiplet part of the effective action was proved
to coincide with the expression found earlier in \cite{BI} by invoking
the requirement of hidden $\N=2$ supersymmetry which completes the manifest $\N=2$
one to the full $\N=4$ supersymmetry. Thus the direct quantum computation reproduces the effective
Lagrangian found in \cite{BI}. Also, we demonstrated that the same $\N=2$ covariant
harmonic supergraph techniques allow one to derive the non-holomorphic $W, \bar W$
potential. We conclude that our approach sets up a generic manifestly
$\N=2$ supersymmetric framework for analysing
the dependence of the full low-energy effective action of $\N =4$ SYM theory
on all fields of ${\cal N}=4$ gauge multiplet. The results of \cite{BI} and this paper
can be regarded as providing the ultimate solution to the problem of constructing the
manifestly $\N=2$ supersymmetric exact low-energy effective
action in $\N=4$ SYM theory. Whereas we considered here the
simplest case of the gauge group $SU(2)$, it is straightforward to extend
our study to the general case of gauge group $SU(N)$ broken down to $U(1)^{N-1}$.

It would be interesting to find, using both the quantum and
algebraic methods, the full $\N=4$ supersymmetric form of some subleading
$W, \bar W$ terms in the effective action of $\N=4$ SYM theory, e.g.
of those studied in \cite{bt,bt2}. This could provide further checks of the
supergravity/super Yang-Mills correspondence and be of direct relevance
to the closely related problem of constructing $\N=4$ superconformally invariant
extension of the Dirac-Born-Infeld theory in $\N=2$ superfield
formulation \cite{iva}.

\section*{Acknowledgements} We are very grateful to S.M. Kuzenko for
critical comments on section 4 of the paper. 
The work of I.L.B. and E.A.I. was partially
supported by INTAS grant, project No 00-00254, DFG grant, project No
436 RUS 113/669 and RFBR-DFG grant, project No 02-02-04002.
The work of E.A.I. was partially supported by the RFBR-CNRS grant,
project No 01-02-22005. A.Yu.P. is grateful
to FAPESP grant, project No 00/12671-7, for support. I.L.B. thanks 
BLTP, JINR and, personally, its Director Prof. A.T. Filippov for kind 
hospitality in Dubna at the final stage of this work.
He is also very grateful to Prof. S.J. Gates for kind hospitality in
the University of Maryland in Fall of 2002. This work has been finalized when 
E.A.I. visited the Institute of Theoretical Physics of the University of Hannover. 
He thanks Profs. O. Lechtenfeld and N. Dragon for kind hospitality.

\end{document}